\documentclass[12pt]{article}
\usepackage{a4,graphicx,fancyhdr, vmargin, psfig, texstyle}
\usepackage[authoryear]{natbib}
\usepackage{amssymb, amsmath,amsthm,graphicx}
\usepackage{tikz}
\usepackage{pgflibraryarrows}
\usepackage{pgflibrarysnakes}
\usepackage{tabularx,gensymb}

\usepackage{hyperref}
\usepackage{changes}
\newcolumntype{R}[1]{>{\raggedleft\arraybackslash}p{#1}}
\renewcommand{\baselinestretch}{2}
\allowdisplaybreaks

\begin{document}

\begin{flushleft}
  {\bf Synthesizing geocodes to facilitate access to detailed geographical information in large-scale administrative data}
  \vspace{1.0cm}

  J{\"o}rg Drechsler$^*$, Jingchen Hu$^{**}$

{\small
\begin{description}
\item $^* \;$ Institute for Employment Research, Regensburger Str. 104, 90478 Nuremberg, Germany, joerg.drechsler@iab.de
\item $^{**}\;$ Department of Mathematics and Statistics, Vassar College, Poughkeepsie, NY 12604, USA, jihu@vassar.edu
\end{description}
}
\end{flushleft}
\vspace{1cm}

\noindent {\bf Abstract}.
{We investigate whether generating synthetic data can be a viable strategy for providing access to detailed geocoding information for external researchers, without compromising the confidentiality of the units included in the database. Our work was motivated by a recent project at the Institute for Employment Research (IAB) in Germany that linked exact geocodes to the Integrated Employment Biographies (IEB), a large administrative database containing several million records.  
We evaluate the performance of three synthesizers regarding the trade-off between preserving analytical validity and limiting disclosure risks: One synthesizer employs Dirichlet Process mixtures of products of multinomials (DPMPM), while the other two use different versions of Classification and Regression Trees (CART). In terms of preserving analytical validity, our proposed synthesis strategy for geocodes based on categorical CART models outperforms the other two. If the risks of the synthetic data generated by the categorical CART synthesizer are deemed too high, we demonstrate that synthesizing additional variables is the preferred strategy to address the risk-utility trade-off in practice, compared to limiting the size of the regression trees or relying on the strategy of providing geographical information only on an aggregated level. We also propose strategies for making the synthesizers scalable for large files, present analytical validity measures and disclosure risk measures for the generated data, and provide general recommendations for statistical agencies considering the synthetic data approach for disseminating detailed geographical information. 
}

\section{Introduction}
In recent years, an increasing number of statistical agencies have started collecting detailed geocoding information on respondents for some databases
. Access to detailed geocoding information empowers researchers to define their own geographical areas of interest by aggregating over the individual geocodes, instead of depending on pre-specified administrative geographical levels such as municipalities or counties when analyzing spatial effects. Furthermore, the 
geocoding information can  be used to facilitate the linkage of data from different sources. However, these additional research opportunities come at a price: the detailed geocoding information makes it very easy to identify individuals in the database, which is why external researchers usually cannot obtain access to the detailed geocodes.

Given the high risks of re-identification if detailed geocoding information is released, most of the traditional methods of statistical disclosure limitation \citep{Hundepool2012book,duncan:2011}, such as swapping or noise infusion, would either have to be applied extensively, severely distorting the relationships between the variables, or would not protect the data sufficiently. While methods offering formal privacy guarantees such as differential privacy \citep{Dwork2006DP} could be applied to ensure a high level of data protection, as recently illustrated by \citet{abowd2018} and \citet{bowen2019}, the amount of noise that would be necessary to allow for a data release on the level of geographical detail considered in our work would be so high, that only very limited information could be extracted from the protected data. A promising alternative in our situation is the release of synthetic data without formal privacy guarantees: Sensitive records or records that have a high risk of disclosure are replaced with draws from a model fitted to the original data. If the synthesis models are carefully developed, important features of the data could be maintained while disclosure risks can be substantially reduced \citep{Drechsler2011}. We refer to \citet{barrientos2018} for an interesting application which combines data synthesis with a formally private verification server.  

In this work, we evaluate three synthesizers for generating synthetic geocodes
. The first approach concatenates the information on latitude and longitude for each record, treating the resulting variable as an unordered categorical variable, and uses Dirichlet process mixtures of products of multinomials (DPMPM) for synthesis. The second approach generates synthetic values by using CART models \citep{Reiter2005CART} for the categorical geocoding variable. The third approach is also based on CART models, but treats the information on the latitude and longitude as two separate continuous variables \citep{WangReiter2012}. We selected these three synthesizers given their promising results in various applications: 
The usefulness of the DPMPM approach was first illustrated in \citet{HuReiterWang2014} with an application to synthesizing a subset of the 2012 American Community Survey (ACS)
; CART models have repeatedly been shown to outperform standard parametric models and other machine learning techniques such as random forests or support vector machines in the imputation/synthesis context \citep{DrechslerReiter2011, drechsler:2012, doove:2014, akande:2017} and their usefulness for synthesizing geocoding information has been demonstrated in \citet{WangReiter2012}. 

Our work was motivated by a recent project at the Institute for Employment Research (IAB) in Germany that added detailed geographical information to the Integrated Employment Biographies (IEB), a rich administrative data source. The IAB is investigating strategies for providing data access for external researchers. The results presented in this work are part of this endeavor, and all
evaluations of the synthesizers are conducted using a subset of variables from these data.

Our extensive evaluation studies suggest that the categorical CART synthesizer outperforms the other two in terms of preserving the analytical validity. However, disclosure risks are substantially higher.
 Assuming the data disseminating agency is not satisfied with the level of protection, we evaluate three strategies for further reducing the disclosure risks when using categorical CART synthesizers: i) aggregating the geocoding information before synthesis; ii) synthesizing additional variables in the dataset; and iii) modifying the complexity parameter -- a tuning parameter in CART models. We show that synthesizing additional variables in the dataset is the preferred strategy for addressing the risk-utility trade-off in practice. We summarize our findings and leave the detailed results in the online supplement.

Several strategies have been proposed in the literature for synthesizing data containing detailed geographical information. However, most of them are not suitable for our application. \citet{abo:ger:08} proposed methodology which offers formal privacy guarantees -- $\varepsilon-\delta$ differential privacy. Their methods require special techniques for dealing with the sparsity of the matrix when cross-classifying two variables, which does not scale up when more variables are involved. \citet{Sakshaug2010} proposed mixed effects models for preserving the geographical information in the synthetic data when synthesizing other non-geographical variables, but not the geographical information itself. \citet{quick2018} also only synthesized non-geographical information using a differential smoothing synthesizer. \citet{paiva2014} proposed areal-level spatial models to synthesize geographical information, which are computationally feasible with only a small number of categorical variables and limited numbers of levels. \citet{quick2015} proposed marked point process models for synthesizing geographical information. The authors pointed out that the model estimation can be computationally intractable and suggested several simplifications for their application of a  a relatively small dataset.
%


The remainder of the paper is organized as follows: 
Section \ref{data} introduces the IEB data. In Section \ref{syn} we describe the DPMPM synthesizer, the categorical CART synthesizer, and the continuous CART synthesizer. Section \ref{syndata} provides some details on the implementation of the synthesizers for the IEB synthesis. In Section \ref{measures_and_results} the analytical validity and disclosure risks of the generated synthetic datasets are evaluated under various assumptions. We conclude with a general discussion of the three synthesizers, implications for the dissemination project at the IAB, our list of recommendations for statistical agencies considering the use of synthetic data to disseminate detailed geocoding information, and some ideas for future research.
\section{The 
Integrated Employment Biographies (IEB)}\label{data}
The IEB integrate five different sources of information collected by the Federal Employment Agency in Germany through different administrative procedures: the Employment History, the Benefit Recipient History, the Participants-in-Measures History, the Unemployment Benefit II Recipient History, and the Jobseeker History. A more detailed description of the data and of the current access procedures for external researchers can be found in the online supplement.

Recently, exact geocoding information has been added for individuals and establishments included in the IEB on the reference date June 30, 2009. The geocoding information was obtained from a georeferenced address database for Germany provided by the Federal Agency for Cartography and Geodesy. This database contains approximately 22 million addresses of German buildings and their corresponding geographic coordinates. Based on exact matches regarding the address information, it was possible to obtain geocoding information for 94.6\% of the 36.2 million individuals and 93.2\% of the 2.5 million establishments contained in the IEB on the reference date (see \citet{ScholzRauscher2012} for 
further details
).
Given the large number of variables and the sensitivity of the information contained, it is obvious that the linked data cannot be disseminated to the public using traditional statistical disclosure limitation techniques
, because the geocoding information would make it very easy to identify individuals in the database. In fact, the data currently available for external researchers only contain a 2\% sample from the IEB with a limited set of non-sensitive variables and county-level information as the lowest level of geographical detail (see the online supplement for 
further details). 

To be able to provide access to the detailed geocoding information for external researchers, the IAB is looking for innovative ways to generate a sufficiently protected version of the linked data that contains useful information, at least on some of the variables. 
The decision was made to start with a limited set of variables initially and extend the set in the future if reasonable results could be achieved in the first round, both in terms of disclosure risk and data utility.
The selected variables are listed in Table~\ref{vars}. In the application, we use an additional variable, ZIP code, which is an aggregation of the exact geocoding information. The ZIP code is commonly used as the lowest level of geographical detail when analyzing spatial effects. We do not use the ZIP code information when synthesizing the geocodes. Instead, synthetic ZIP codes are derived directly from the synthetic geocodes (see Section \ref{utility_results} for details).

\renewcommand{\baselinestretch}{1}
\begin{table}[t]
\centering
\begin{tabular}{ll}
    \noalign{\smallskip}
variable & characteristics\\
   \hline\noalign{\smallskip}
exact geocoding information 	&recorded as distance in meters from the point\\
of the place of residence          & 52\degree north, 10\degree east \\
sex &	male/female\\
foreign 	  &	yes/no\\
age &	6 categories \\
&($<$20, 20--30, 30--40, 40--50, 50--60, $>$60)\\
education   &	6 categories\\
occupation level &	7 categories\\
occupation&	12 categories\\
industry of the employer &	15 categories\\
wage  &	10 categories defined by quantiles\\
distance to work&5 categories ($\leq$1, 1--5, 5--10, 10--20, $>$20 km)\\
   \hline\noalign{\smallskip}
\end{tabular}
\caption{Variables included in the data set used for the evaluations}\label{vars}
\end{table}
\renewcommand{\baselinestretch}{2}

\section{Modeling assumptions for the different synthesizers}\label{syn}
In this section, we briefly describe the DPMPM, the categorical CART, and the continuous CART synthesizers for generating synthetic data
. More details on three synthesizers can be found in Appendices \ref{DPMPM_detail} and \ref{CART_detail}.

The DPMPM is a Bayesian semi-parametric procedure. It uses a Dirichlet process mixture of products of multinomial distributions, which is a Bayesian latent class model. The basic idea is to assume that each observation belongs to one of a potentially infinite number of latent classes. Conditional on the latent class assignment, the variables contained in the data are considered independent. It can be shown that the DPMPM provides full support on the space of distributions for multiple unordered categorical variables \citep{dunsonxing}. See Appendix \ref{DPMPM_detail} for more details on how the DPMPM can be turned into an engine for data synthesis.

The CART approach is nonparametric and is based on classification and regression trees from the machine learning literature. As outlined in \citet{DrechslerReiter2011}, the approach seeks to approximate the conditional distribution of a univariate outcome from multiple predictors. The CART algorithm partitions the predictor space so that subsets of units formed by the partitions have relatively homogeneous outcomes. The partitions are found by recursive binary splits of the predictors. The series of splits can be effectively represented by a tree structure, with leaves corresponding to the subsets of units. The values in each leaf represent the conditional distribution of the outcome for units in the data with predictors that satisfy the partitioning criteria that define the leaf. CART has been adapted for generating partially synthetic data \citep{Reiter2005CART} and we refer to Appendix \ref{CART_detail} for more details on how this can be achieved. In our evaluations, we use a categorical CART synthesizer and a continuous CART synthesizer.

In this article, we assume that the aim is to generate partially synthetic data, i.e., only some of the variables in the released data will be synthesized. We note that DPMPM and CART synthesizers can also be used to synthesize all variables, i.e., to generate fully synthetic data. 

\section{Synthesis of the IEB}\label{syndata}
For our evaluations, we selected individuals living in Bavaria and deleted all observations with missing information in any of the variables. Since most of the variables (such as wage or distance to work) are only observed if the individual is employed on the reference date, the final dataset, consisting of 3,333,998 records, represents the working population in Bavaria. In practice, it would not be necessary to remove the records with missing values. 
Since all variables are categorical, missing values could simply be treated as another category during the synthesis. Alternatively, all missing values could be imputed first and then the imputed data could be synthesized following the two-stage process proposed in \citet{reiter:2004}. We deliberately decided to remove all cases with missing values, since this study aims at evaluating the performance of different synthesizers when dealing with geocoding information. If all records were included, it would be impossible to disentangle the performance of the methods when dealing with missing values from the performance of the methods when synthesizing geocoding information. 

 We initially only synthesized the geocoding information for the place of residence. If the disclosure risks based on this strategy are deemed too high, it is straightforward to use the synthesis models described above to synthesize additional variables in the dataset. We explore the impacts of additional synthesis on risk and utility in the online supplement, and summarize our findings in Section \ref{do_more}.

Fitting one synthesis model to the entire dataset would be prohibitive due to the size of the data, therefore we clustered all of the observations into 222 similar-sized clusters based on their geographic locations, each containing 15,000 records (except for the last cluster, which contains 18,998 records), and ran the synthesis models separately for each cluster. We achieved the clustering using the maximum distance to average record (MDAV) algorithm \citep{DomingoMateo2012}. Since the synthesis models are independent between the clusters, it is possible to run the synthesis for each cluster in parallel. In this way, the synthesis procedure becomes scalable even if the entire dataset for Germany -- with more than 36 million records -- should be synthesized. We note that running the synthesis only within clusters results in increased risks of disclosure, because the outcome space of the synthetic geocodes is bounded by the geocodes observed within the cluster. We take this into account when evaluating disclosure risks in Section \ref{risk}.

In each cluster, we ran the MCMC sampler for the DPMPM synthesizer for 10,000 iterations, treating the first 5,000 iterations as burn-in and storing only every 10th iteration to reduce the correlation between the successive draws. Necessary convergence diagnostics were performed, and the details are included in the online supplement.

For the categorical CART synthesizer, we used the default values implemented in the $rpart$ package in R \citep{rpart} for two of the three tuning parameters that control the size of the trees that are grown: i) the default for the minimum number of observations that must exist in a node in order for a split to be attempted is 20; ii) the default for the minimum number of observations in any terminal leaf is seven. However, we set the the third tuning parameter -- the complexity parameter -- to a very low value of $cp=0.00001$ (the default is $cp=0.01$). Any split that does not decrease the overall lack of fit by a factor of $cp$ is not attempted. Since we are interested in preserving the relationships between the geocodes and the other variables in the dataset as closely as possible to obtain high analytical validity of the synthetic data, we choose a very low level for this parameter. We note that this parameter is the most powerful of the three tuning parameters for balancing the analytical validity and the disclosure risk of the generated data. If the risks are considered too high, the synthesis could be repeated using a larger $cp$ value. We explore changing the value of the complexity parameter in the online supplement and summarize our findings in Section \ref{do_more}. 

For the continuous CART synthesizer, we only report the results for the longitude-then-latitude synthesis, as our evaluation indicated that the reverse order, latitude-then-longitude synthesis, produced almost identical results
.

We generated $m=5$ synthetic datasets for each cluster for all synthesizers. We note that releasing only one synthetic dataset is recommended if the data are only used to develop software code to be run on the original data, especially as increasing $m$ will typically lead to an increased risk of disclosure \citep{DrechslerReiter2009,ReiterDrechsler2010}. However, multiple synthetic datasets will reduce the Monte Carlo error and are generally required to obtain valid variance estimates based on the synthetic data \citep{reiter:2003}.

\section{Analytical validity and disclosure risk}\label{measures_and_results}
In this section, we evaluate the performance of the three synthesizers in terms of addressing the trade-off between preserving the analytical validity (Section \ref{utility}) and offering a sufficient level of data protection (Section \ref{risk}). We start each section by presenting first the evaluation criteria, then the results for the synthetic IEB data.  We also discuss strategies for further increasing the level of protection in Section \ref{do_more}.

\subsection{Analytical validity evaluation}\label{utility}


Two types of validity measures are typically distinguished in the literature: i) global utility measures, which assess the validity of the generated data by computing some distance measure between the original and the protected data; and ii) outcome-specific utility measures, which evaluate the extent to which 
the results of a specific analysis are preserved. Since both measures have advantages and disadvantages
, we aimed at assessing both dimensions of analytical validity.


\subsubsection{Measures for evaluating the analytical validity}\label{utility_measures}

\noindent We considered several outcome-specific utility measures. The first set of outcome-specific utility measures assumes that the analyst 
analyzes regional differences in the population at a very detailed geographical level, such as the ZIP code level. 
With access to the exact geocoding information, the analyst is very flexible in defining the geographical area she is interested in. For our evaluation, we choose the ZIP code level for three reasons: i) we need a sensible strategy to cluster all of the data for Bavaria, and the ZIP code offers a natural way of clustering the data on a very detailed level without the need to specify arbitrary boundaries for setting up the clusters; ii) the number of records vary considerably between ZIP codes, ranging from just a few cases in rural areas to more than 10,000 cases in densely populated areas (the median size being 535 cases), and therefore, the ZIP code level evaluation covers both statistics which should be relatively easy to preserve (since they are based on a large number of cases) and those for rural areas, which will be difficult to preserve; and iii) choosing the ZIP code level offers the convenience that maps can be generated relatively easily using any GIS software, since the geocodes of the ZIP code boundaries are directly available.

We recommend the use of heat maps to visualize the distribution of the outcome of interest for the original data and for the synthetic data generated by different synthesizers. Graphical comparisons of the heat maps help to assess the level of analytical validity of synthetic data.

The other set of outcome-specific utility measures includes Ripley's $K$- 
and $L$-functions \citep{Ripley1976, Ripley1977}
, two commonly used descriptive statistics for detecting deviations from spatial homogeneity. Since the data we use in our evaluations only contains categorical variables, any outcome of interest will have a multitype point pattern, that is, a pattern in which each point is classified as belonging to one of a finite number of possible types \citep{spatstatJSS}. 
We therefore consider the multitype $K$-function, which counts the expected number, $\hat{K}_i(r)$, of other data points within a given distance, $r$, of a point of type $i$ \citep{LotwickSilverman1982}: 
\begin{equation*}
\hat{K}_i(r) = |D_s|\sum_{i \neq j} \frac{I (||{\bf{s}}_i - {\bf{s}}_j|| \leq r)}{n\times n_i},
\end{equation*}
where $\bf{s}$ is the spatial domain, $|D_s|$ is the area of the spatial domain, $n$ is the total number of points, and $n_i$ is the number of points of type $i$. For example,
one can calculate the value of $\hat{K}_i(r)$ for individuals aged 60 and older (type $i$) 
within 100 meters ($r$) in a particulay city ($D_s$). In this example, $\bf{s}$ contains the geocoding information for all individuals living in the selected city, $n$ is the number of individuals living in the city, and $n_i$ is the number of individuals in the city who are 60 or older. 

To measure spatial dependence, it is common to compute the $L$-function \citep{quick2015}
\begin{equation*}
\hat{L}_i(r) = \sqrt{\hat{K}_i(r)/\pi}- r.
\end{equation*}
For approximately homogeneous data, the expected value of $\hat{L}_i(r)$ is zero, and positive values of $\hat{L}_i(r)$ indicate spatial clustering. 


Common measures of global utility are the Kullback-Leibler divergence or Hellinger distance. A downside of these distance measures is that they return only one value to assess the overall validity, potentially losing important information. 
To overcome these shortcomings, we propose a simple approach for evaluating the global utility which shares some similarities with the measures proposed by \citet{shlomo:2010}, \citet{Woo2009JPC}, and \citet{Raab2017arXiv} to evaluate the utility for frequency tables:
we compare relative frequencies for various cross tabulations of the variables contained in the original data and the synthetic data.



\subsubsection{Results of the analytical validity evaluations}\label{utility_results}

\begin{figure}[t]
\centering
  \includegraphics[scale=0.46,trim=1cm 1cm 1cm 1cm]{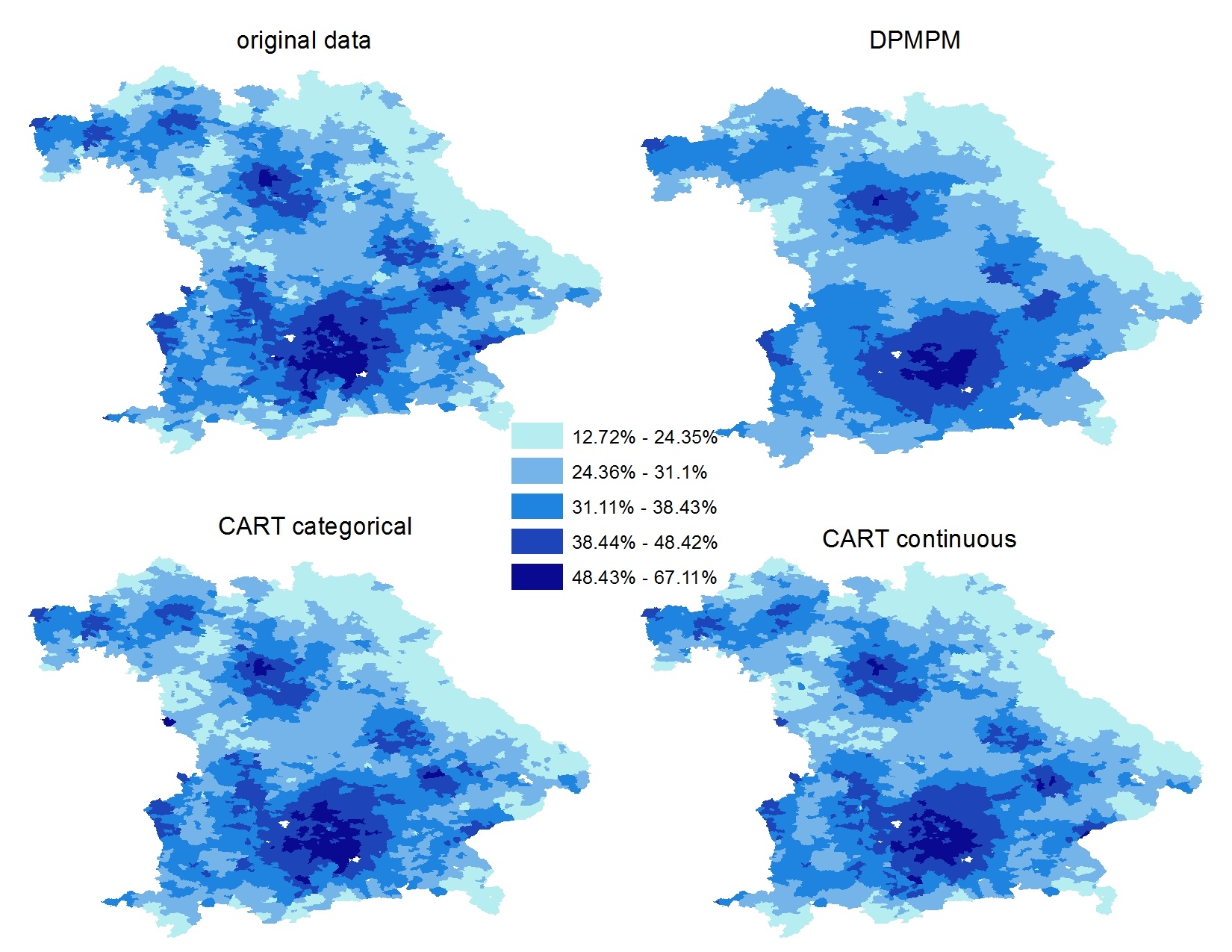}\\
  \caption{Share of high wage earners in Bavaria by ZIP code level.}\label{fig:wage}
\end{figure}
\begin{figure}[t]
\centering
  \includegraphics[scale=0.46,trim=1cm 1cm 1cm 1cm]{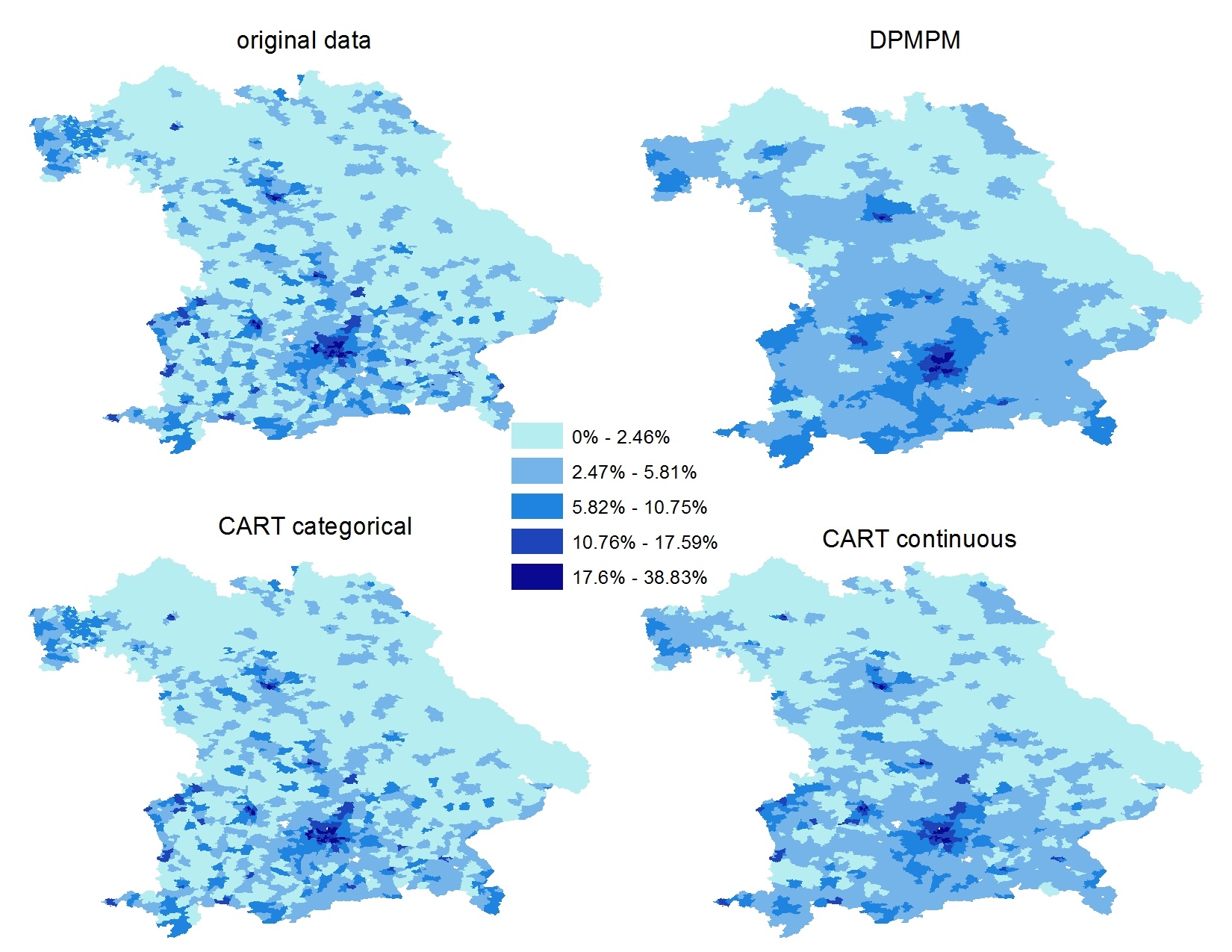}\\
  \caption{Share of foreigners in Bavaria by ZIP code level.}\label{fig:frgn}
\end{figure}

Figure \ref{fig:wage} depicts the share of high wage earners in Bavaria at the ZIP code level using heat maps, where high wages are defined as wages above the seventieth quantile of the wage distribution in Bavaria. We compute the ZIP codes for the synthetic data by identifying the closest geocode in the original data and transferring its ZIP code information. Figure \ref{fig:frgn} depicts the share of foreigners at the same level of geographical detail. Both figures display the results based on the original data and the results based on the three different synthesizers.

All synthesizers preserve the wage distribution rather well. The results are more diverse for the foreigners. This is not surprising given that 30\% of the records in the data are high wage earners by definition, whereas only 6.17\% of the individuals are foreigners, hence the distribution is 
more skewed and thus more difficult to preserve.

For both figures, it is evident that the DPMPM synthesizer does not preserve the geographical heterogeneity as well as the CART synthesizers. The DPMPM model seems to smooth the distributions so that the shares are overestimated in those regions where the shares are low in the original data. This is especially evident in Figure \ref{fig:frgn}. 

The two CART models preserve the wage distribution similarly well. However, the continuous CART model fails to preserve the distribution of foreigners. As with the DPMPM models, the distribution is smoothed out compared to the distribution in the original data, albeit to a lesser extent
. For the categorical CART model, we do not find any substantial differences between the original data and the synthetic data, indicating a very high level of analytical validity.


\begin{figure}[t]
\centering
\includegraphics[scale=0.43]{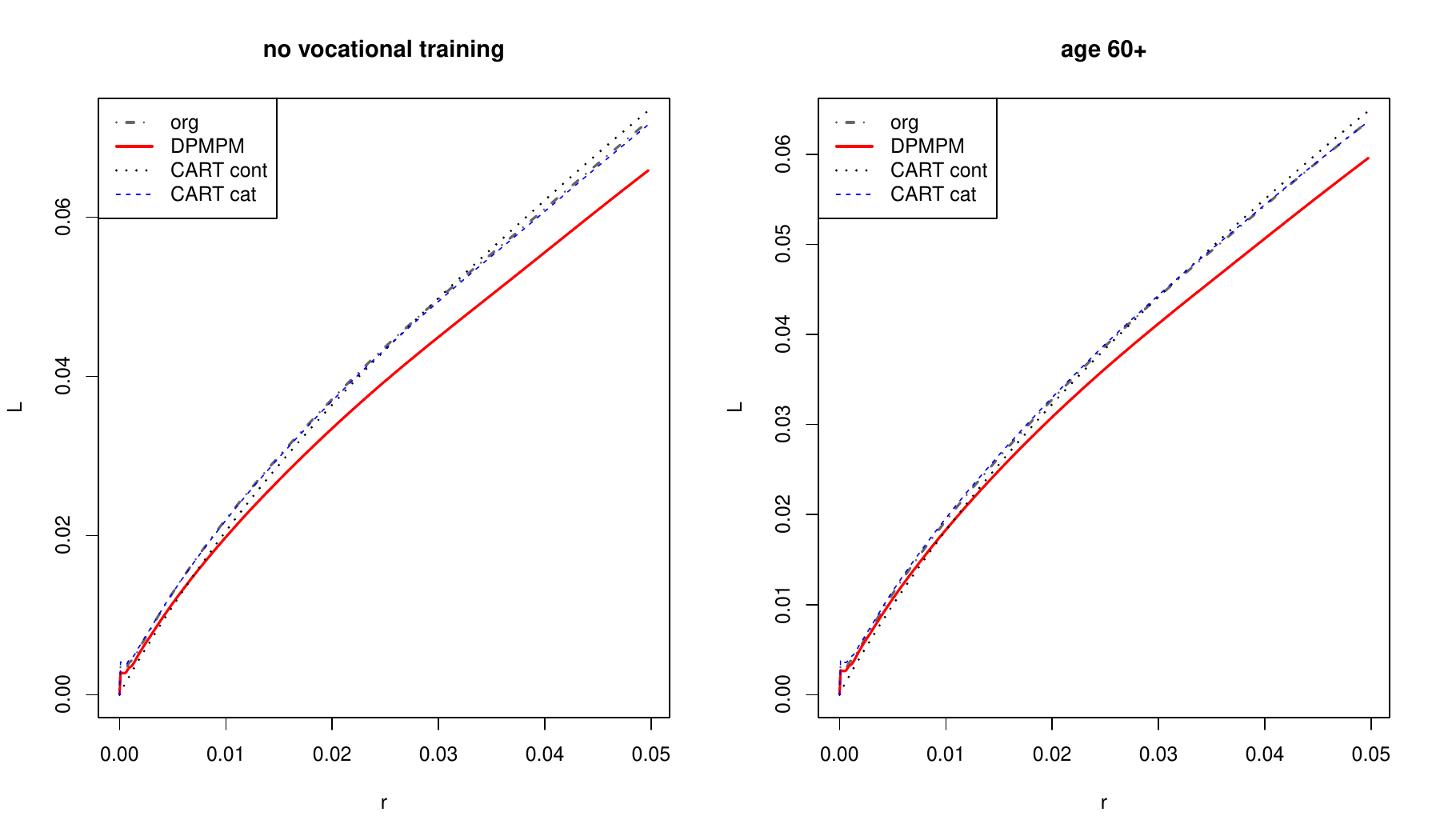}
\vspace{-0.8cm}
  \caption{$L$ functions for individuals without vocational training (left) and individuals aged 60 or above in the city of Nuremberg (right). Each figure contains the results for the original data and the three different synthesizers.}\label{fig:lfunction}
\end{figure}

For outcome-specific utility evaluation focusing on spatial clustering, we use the R package $spatstat$ \citep{spatstatJSS} to calculate the $\hat{L}_i(r)$ functions for the distribution of i) individuals with no vocational training, and ii) individuals aged 60 and above in Nuremberg, Bavaria. For each synthesizer, we calculate the $\hat{L}_i(r)$ functions for the same range of $r$ values in each of the synthetic datasets and average them across datasets. Figure \ref{fig:lfunction} compares the resulting curves to the curve obtained using the original data.

For the range of $r$ values, all synthesizers preserve the degree of clustering in the original data rather well. The difference between the estimated values of $L$ from the original data and the synthetic data is small for both variables for all levels of $r$ considered here. Still, the DPMPM results diverge most from the 
results based on the original data, especially for larger values of $r$. The continuous CART synthesizer offers some improvements, while for the categorical CART synthesizer the results are basically indistinguishable from those 
based on the original data. The results for Nuremberg are in line with the findings for Bavaria discussed above: in terms of preserving the analytical validity, the categorical CART synthesizer should always be preferred, while the DPMPM synthesizer provides the least favorable results.


To evaluate the global utility, we compare the relative frequencies of various cross tabulations of the data. For data including geographical information, as in the IEB data, 
the level of geographical detail is left to the discretion of the analyst. For our evaluation, we choose the ZIP code as the level of geographical detail for the reasons laid out in Section \ref{utility_measures}. Specifically, for each ZIP code we first compute the relative frequencies for each cell entry for various cross classifications of all variables, that is, all marginal distributions, two-way interactions, and three-way interactions. Then we evaluate how much these relative frequencies differ between the original data and the synthetic data. 

\begin{figure}[t]
\centering
  \includegraphics[scale=0.67,trim=1cm 1cm 1cm 1cm]{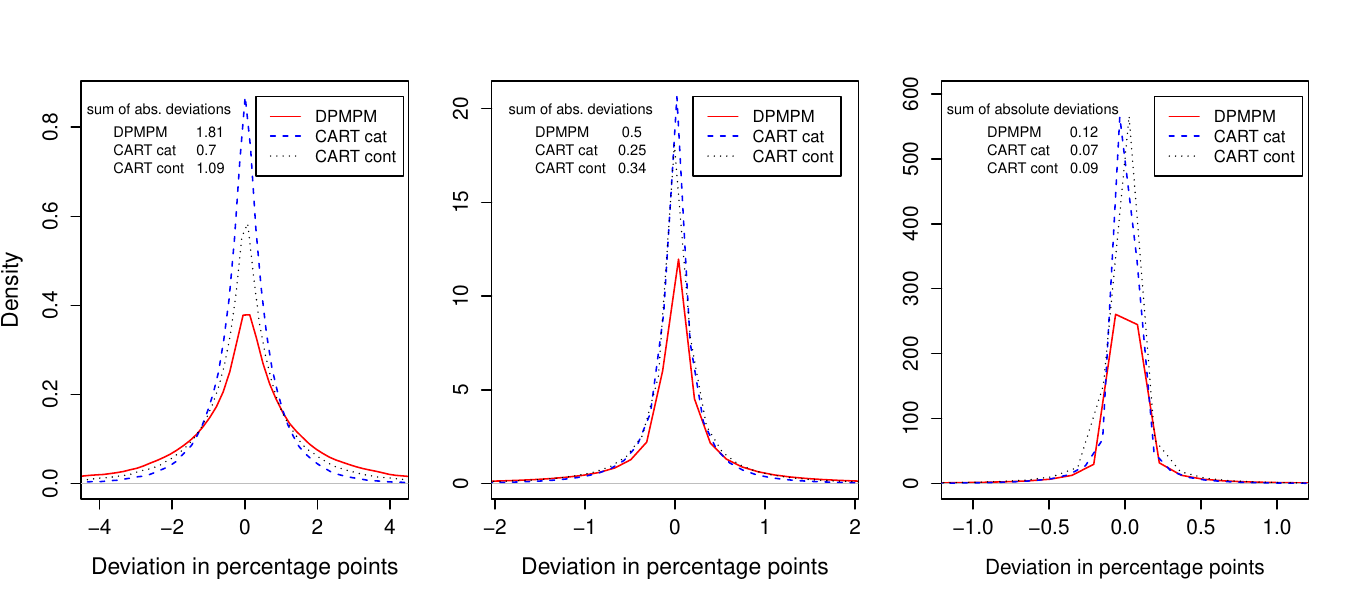}\\
  \caption{Global utility results. Distributions of differences in relative frequencies between the original data and the protected data for various tabulations. The three panels depict the results for all oneway tables (left panel), all twoway tables (middle panel), and all threeway tables (right panel). The numbers in the upper left corner in each panel represent the average absolute values of the differences for the different synthesizers.}
  \label{fig:delta}
\end{figure}

Figure \ref{fig:delta} contains the distribution of the differences across all cells for different interaction levels. The left panel contains all marginal distributions, two-way interactions are presented in the middle panel, while the right panel contains the results for all three-way interactions. Since the utility measure effectively measures the difference in relative frequencies between the original data and the protected data, the higher the density of the distribution around zero, the better the analytical validity. To get a one-number measure for the loss in data utility, we also compute the average absolute values of the differences across all cells (we term this measure the \textit{UL} measure). The numbers for the \textit{UL} measure are reported 
in the upper-left corner of each panel in Figure \ref{fig:delta}. The lower the number, the better the utility. 

The results are similar to those regarding the outcome-specific utility measures: the synthesizer based on the categorical CART model provides the best result, followed by the CART model that treats the geocodes as continuous variables. The DPMPM model consistently performs the worst. Looking at the one-number measure, the differences are quite dramatic: the values of the DPMPM model are always more than 70\% higher than the values for the categorical CART model.

\subsection{Disclosure risk evaluation}\label{risk}

\subsubsection{Measures for evaluating the disclosure risk}\label{risk_measures}

The results from the previous section indicate that from a utility perspective the categorical CART model should always be preferred. However, when disseminating confidential data to the public, there is always a trade-off between data utility and disclosure risk. The categorical CART synthesizer will only be useful if the increase in the risk of disclosure that results from the increase in utility is deemed acceptable. Thus, we will compare risks of re-identification for the various synthesizers in this section. While this will be useful for comparing the different synthesizers in terms of their protection levels, we note that risks of attribute disclosure would also need to be evaluated before any actual data release. 

To evaluate re-identification risks, we propose computing probabilities of identification using methods developed in \citet{ReiterMitra2009}. A detailed description of the methodology is given in Appendix \ref{risk_detail}. Here, we only summarize the main ideas. Suppose the intruder has information on some target records which she will use in a record linkage attack to identify the targets in the released data. Similar to the concept of probabilistic record linkage, the idea is to estimate the probability of a match between each target record and each record in the released file. The record with highest average matching probability across the synthetic datasets is the declared match. In the evaluation, we use two risk measures that are summaries of these matching probabilities: the expected match risk, which computes the expected number of correctly declared matches, and the true match rate, which computes the number of correct single matches among the target records.

We note that the proposed risk measures assume that the intruder knows that the target records she is looking for are included in the database. This is reasonable to assume in our application, since the IEB covers the entire population. If the IAB decides to release only a sample of the synthetic IEB to increase the confidentiality protection further, the risk measures could easily be adjusted to account for the sampling uncertainty using the extensions proposed in \citet{dre:rei:08}.

\subsubsection{Results of the disclosure risk evaluations}\label{risk_results}

For our disclosure risk evaluations, we assume that the intruder knows the exact geocode, sex, age category, industry of the employer, occupation, and the information on whether the individual is a foreigner, and uses this information to try to identify the individuals in the database. If the intruder successfully identifies a single unit in the data, she would learn sensitive information, such as the wage level of the identified individual. We sample 100 records from each cluster and assume that these 22,200 records are the target records that the intruder tries to find in the data. Since the geocodes have been synthesized, only declaring a match if the original and synthetic geocode match exactly might not be the optimal strategy for the geocodes. However, if there are neighborhood effects in the data, the synthetic geocode is still likely to be close to the original geocode. In this case, the intruder 
might still be able to identify the correct record by using caliper matching, where each record that belongs to the same geographical area is considered a potential match. To account for this, we assume that the intruder constructs grids of different sizes and considers all records that fall in the same grid to be matches. We evaluate the risks for four different grids: 100$\times$100, 1,000$\times$1,000, 10,000$\times$10,000, and 20,000$\times$20,000 square meter grids. We also compute the risk measures if the intruder matched on the exact geocodes. To account for the increased risk from clustering the data before the synthesis, we block on the cluster when matching the target records. This is a conservative approach, since the intruder would not be able to identify the clusters in the released data. Still, we believe most agencies would prefer being conservative instead of underestimating the risks of a data release strategy.  

The estimated risks for the original data without the geocodes are as follows: expected risk: 1821.16; and true rate: 2.56\%.
These measures serve as a benchmark, since under the current regulations external researchers can already access a 2\% sample of the IEB (without geocodes) through the Research Data Center of the IAB. So any synthetic data with similar risks should be considered sufficiently protected as long as the data can only be accessed under similar conditions.

\begin{figure}[t]
\centering
  \includegraphics[scale=0.6]{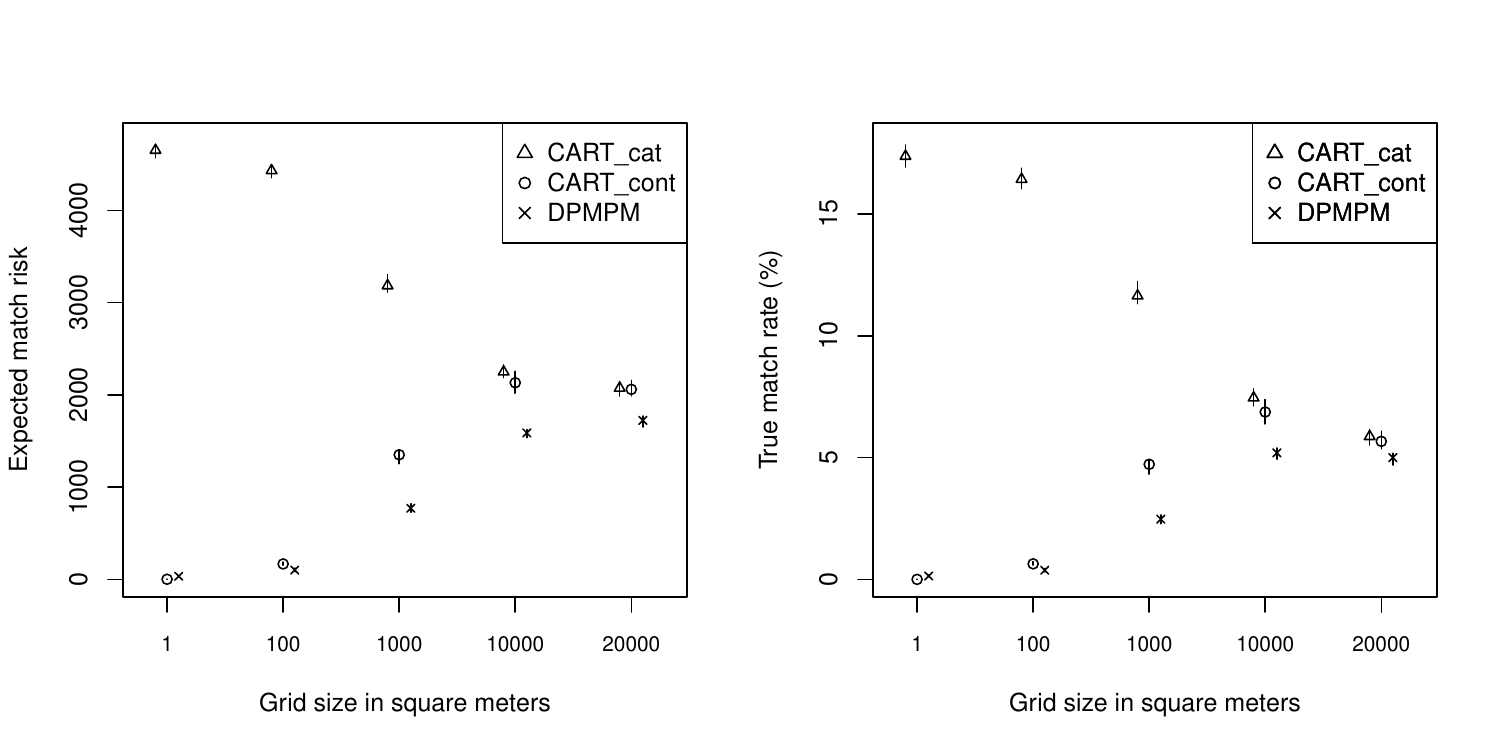}\\
  \caption{Expected match risk and true match rate (in \%) for various grid sizes. The different symbols represent the results for the different synthesizers averaged over the ten simulation runs. The vertical lines stretch from the smallest to the highest value observed across the ten simulation runs.}
  \label{fig:risk}
\end{figure}

To ensure that the risk assessments can be generalized, we repeat the risk evaluations for ten different sets of randomly sampled target records and assess the variability across the different runs. The results are presented in Figure \ref{fig:risk}. The different symbols represent the average results, and the end points of the vertical bars are the minimum and maximum values observed across the ten runs. As expected, the risks increase with increasing analytical validity. The DPMPM synthesizer that showed the lowest analytical validity is associated with lowest risk of disclosure for both matching strategies in almost all scenarios.
The categorical CART synthesizer, which has very high analytical validity, also leads to the highest risk of disclosure.

For the DPMPM and the continuous CART synthesizer, the risks generally increase with increasing grid size
. Arguably, with these two synthesizers the risks do not increase substantially compared to the risks for the data without geocodes. The expected risks are slightly larger for the continuous CART if the intruder uses grids of 10,000 square meters for the matching, with an average risk of 2132.13 compared to 1821.16 for the data without geocodes. The percentage of correct unique matches (the true match rate) is also larger for both the continuous CART and the DPMPM if the intruder chooses grid sizes $\geq$1,000 square meters with a maximum of 10,000 square meters for both synthesizers. The average true match rate for this grid size is 6.87\% for the continuous CART and 5.19\% for the DPMPM, compared to 2.56\% for the original data without geocodes. However, these increased risks are balanced by the fact that the intruder would not know which grid size to pick.



The results are different for the categorical CART synthesizer. Here, the risks decrease monotonically with increasing grid size, and both the average expected risk (with a maximum of 4652.78) and the average true match rate (with a maximum of 17.37) indicate a substantial increase in risk compared to releasing the data without geocoding information. In practice of course, an intruder would not know which grid size implies the highest risk of disclosure. Looking at the maximum risk across the different grid sizes thus gives a conservative upper bound for the risk of disclosure.

The diverging results for the different synthesizers can be explained by observing and weighing two competing effects of increasing grid size on whether a record will be declared a match: On the one hand, it will increase the probability of the correct match being included among the declared matches; on the other hand, it will increase the probability of finding more than one matching record, which decreases the probability of the correct match being identified. Given the low utility of the continuous CART and DPMPM synthesizer, the geocodes in the synthetic data are unlikely to be close to the original ones; therefore, the first effect will dominate for these two synthesizers. Given the high utility of the categorical CART synthesizer, the generated synthetic geocode is close to the original geocode for a sufficient number of cases; therefore, the first effect is outweighed by the second effect of the increased probability of finding more records that match on all matching criteria. Interestingly, the risk results converge for the three synthesizers when looking at the largest grid size. We speculate that this is because with such large grid sizes it does not make much difference whether the geocoding information is used during matching or not.

\subsection{Strategies for further increasing the level of protection}\label{do_more}

If the agency feels that the increase in risk with the categorical CART synthesizer is not acceptable, the agency would have three options: i) use one of the other synthesizers and accept the loss in analytical validity, ii) use additional measures to protect the data further, or iii) repeat the synthesis, increasing the complexity parameter to grow smaller trees.
We evaluated the last two options for various settings. For the sake of brevity, we only present the main findings for the categorical CART synthesizer here. Detailed results can be found in the online supplement. 

As additional data protection measures, we examined two different approaches. The first approach aggregates the geographical information to a higher level. The second approach synthesizes additional variables in the dataset. Aggregating the geographical information is a classical data protection measure that is commonly employed by statistical agencies before data dissemination. Thus, we evaluated how the risk-utility profile changes if the detailed geographical information is aggregated to a higher level before the synthesis. Specifically, we looked at the impacts of using 10, 100, and 1,000 square meter grids instead of the exact geocodes. We found that risks remain constant for the 10 square meter grids and actually increase if 100 square meter grids are used. The risks decrease for the 1,000 square meter grids but are still  higher than the risks from the original data without geocodes (for example, the expected risk is 3717.65, compared to 1821.16 for the original data). At the same time, the loss in analytical validity, as expressed by the \textit{UL} measure, would increase for such large grid sizes (the \textit{UL} measure increases from $8.57\cdot10^{-2}$ when synthesizing the exact geocodes to $8.74\cdot10^{-2}$ when using 1,000 square meter grids).

As an alternative, the IAB could decide to synthesize more variables if the level of protection from synthesizing the geocodes is not deemed to be sufficient. We looked at different synthesis strategies, starting with synthesizing one additional variable (age), successively adding occupation and foreign status to the list of synthesized variables. In the final setting, we synthesized all variables except for wage and sex. As expected, both risk and utility will decrease with increasing amounts of synthesis. However, the decreases in risk are more pronounced. While the \textit{UL} measure only increases by 4.1\% to 7.9\% for the first three scenarios, and by 14.8\% for the scenario in which almost all variables are synthesized, the expected match risk decreases by between 60.9\% and 85.6\% for the first three scenarios, and 99.7\% for the final scenario. We also find that when synthesizing almost all variables in the dataset, the analytical validity of the categorical CART synthesizer is still higher than if any of the other synthesizerswere used for synthesizing the geocode alone. Again, we refer to the online supplement for a detailed discussion of the results.

Finally, we evaluated how changing the complexity parameter ($cp$) of the CART model affects the analytical validity and risk of disclosure. Based on the extensive simulation studies described in detail in the online supplement, we find--not too surprisingly--increasing the $cp$ value will decrease both the analytical validity and the risk of disclosure. However, the relationship is best described as a stepwise function. Utility and risk remain more or less constant over large ranges of the $cp$ value, but can sometimes change substantially with small changes of the $cp$ value (see Figure 5 in the online supplement). This effect can be explained by looking at the sizes of the grown trees, which show a similar behavior. The tree sizes remain almost constant for several $cp$ values, but can sometimes change substantially with small changes in the $cp$ value (see Figure 4 in the online supplement). 

Overall, our evaluations indicate that only the strategy of synthesizing more variables seems suitable to address the risk-utility trade-off in practice. Aggregating the geocoding information leads to substantial information loss but only modest gains in disclosure protection. Modifying the $cp$ value will be cumbersome in practice, since it is the $cp$ values
that actually have an impact on risk and utility which need to be identified, and it is not obvious that a value can be identified 
 which achieves the desired balance between risk and analytical validity. This does not mean that practitioners should not evaluate different $cp$ values. We find that changing the $cp$ values can have substantial effects on validity and risk, and thus it is generally recommended to run synthesis models with different $cp$ values. The $cp$ value is just not a useful tuning parameter for finding the optimal value on the risk-utility frontier.

\section{Discussion}\label{discussion}

Access to detailed geographical information is desirable in many situations. However, providing this access is challenging in practice, since the risks of re-identification will increase substantially. Therefore, innovative data protection procedures offering a very high level of protection are required if this information is to be released. In our view, the synthetic data approach is the only viable solution for this endeavor.


In this paper, we used three different synthesizers for synthesizing the geocoding information, and evaluated how well they address the common dilemma in data confidentiality: the trade-off between disclosure risk and analytical validity. The continuous CART synthesizer has been suggested previously as suitable for protecting detailed geocoding information. 
The DPMPM synthesizer recently gained much attention due to its improved performance compared to other parametric approaches for categorical variables. 
However, we found that using CART models and treating the geocodes as categorical variables provided the best results.
The relatively poor performance of the DPMPM might be due to the fact that the DPMPM approach tries to model the full joint distribution of all the variables, although we actually only need a good model for the conditional distribution of the geocodes given all the other variables in our synthesis application. This is exactly what the CART-based approaches try to model, and it might be easier to capture this conditional distribution instead of the full joint distribution. In a comparison of the two CART approaches, the continuous CART synthesizer might be inferior in terms of analytical validity, because geographical proximity might not be a good measure for setting up the splitting rules (with continuous variables, CART will try to minimize the variance in each leaf, which might not be a useful criterion with geocodes).

The continuous CART has the additional disadvantage that it can produce implausible geocodes, such as places of residence in the middle of a lake, in a forest, or in industrial areas. This will not happen with the categorical CART, since only geocodes that were observed in the original data can appear in the synthetic data.
This could be problematic in some applications in which the collected data only comprise a sample of the population and the information that an individual participated in the survey is confidential in itself. Releasing unaltered geocodes would put some individuals at risk even if the synthetic geocodes were attached to different units, since the information would be revealed that someone living at a specific geographical location included in the data must have participated. However, since the IEB covers the entire population, this is less problematic in our context. All individuals from the population should be included in this database, and the information that someone lives at a specific location is generally not confidential.

Nevertheless, 
the increased validity of the categorical CART approach comes at the price of increased risks of disclosure. However, at least in this application, we found that the agency would be better off synthesizing more variables instead of relying on one of the other synthesizers, aggregating the geocodes, or modifying the complexity parameter of the CART models.
We note that our risk evaluations did not consider the fact that samples of the IEB without detailed geographical information had previously been disseminated. Potentially, intruders could use this information when trying to re-identify units in the synthetic data. However, the data are only available to the scientific community and underwent several anonymization procedures. We feel that our assumption that the intruder knows the exact geocode, sex, age category, industry of the employer, occupation, and the information on whether the individual is a foreigner or not, is already a conservative one. The only additional variables available in the previously released data are education and wage, but with the low level of geographical detail in these data, it would not be possible for the intruder to obtain record-level information for these variables for the target records. We would thus expect the increase in risk to be negligible.

Based on our findings, we provide the following general recommendations for statistical agencies that are considering using the synthetic data approach to disseminate detailed geocoding information:
\begin{enumerate}
\item
The categorical CART synthesizer should be used in preference to the DPMPM and the continuous CART. It preserves the analytical validity better than the other two synthesizers considered in this paper.
\item
If the risks based on the synthetic geocodes are deemed too high, the agency should consider synthesizing more variables. Depending on the amount of synthesis, the risk of disclosure is quickly reduced with this strategy, and often with a small sacrifice in regard to analytical validity.
\item
Aggregating the geographical information before synthesis is not recommended. It offers only little additional protection unless the level of aggregation is large. The loss of information will be substantial in this case.
\item
Modifying the $cp$ value in the CART-based synthesizers to find the optimal balance between analytical validity and risk is also not recommended. This practice can be cumbersome, and it is not obvious that it would be possible to find a value to achieve this goal. 
\item
For large databases, clustering the data before the synthesis is recommended. The increases in risk are moderate if sufficiently large clusters are selected, but efficiency gains are possible by parallelizing the synthesis of the clusters.
\end{enumerate}

Regarding the actual implementation of the synthetic data approach for the IEB data, the initial results presented in this paper are encouraging enough to consider the approach as a realistic strategy for disseminating the data. Still, several problems need to be tackled before the approach can be used in practice: First, the set of variables will have to be expanded to satisfy user needs (the original data contain more than 50 variables). In this case, it might no longer be sufficient to protect the geographical information alone. In principle, the synthesis approach could easily be extended by simply synthesizing additional variables, as illustrated in Section \ref{do_more}. 
However, modeling some of the variables included in the IEB, such as dates when each spell of employment begins and ends, is a challenging task. Furthermore, the exact geocodes are not only available for the place of residence but also for the place of work. Providing access to both geocodes would allow many additional research questions to be addressed, for example, regarding commuting behavior. In this case, the synthesis methods would have to be adapted to deal with this additional information. Simply fitting classification trees using the place of residence as an additional predictor when synthesizing the place of work will not be an option given the amount of detail contained in the former. At the same time, releasing both geocodes will substantially increase the risk of disclosure. Addressing these issues will be an interesting area of future research. Whether the synthesizer should be extended first to address all these problems, or whether a reduced subset of variables should be released at this stage, currently remains a matter of debate.

\section*{Acknowledgements}
We are very grateful to the editor, the associate editor, and the referees for their useful comments and suggestions, which helped improve our manuscript.

\bibliographystyle{natbib}
\bibliography{geo_synth_bib}

\appendix

\section{The DPMPM Synthesizer}\label{DPMPM_detail}

Following \citet{HuReiterWang2014}, let the confidential data $\mathbf{D}$ comprise $n$ individuals measured on $p$ categorical variables.
For $i=1, \dots, n$ and $k=1, \dots, p$, let $y_{ik}$ denote the value
of variable $k$ for individual $i$, and let $\mathbf{y}_i = (y_{i1}, \dots, y_{ip})$.
Without loss of generality, assume that each $y_{ik}$
takes on values in $\{1,\dots, d_{k}\}$, where $d_{k} \geq 2$ is the total number of categories for variable $k$.
Effectively, the survey variables form a contingency table of  $d = d_1\times d_2\times\dots\times d_p$ cells defined by cross-classifications
of the $p$ variables. Let $Y_{ik}$ and $\mathbf{Y}_i$ be random variables
defined on the sample spaces for $y_{ik}$ and $\mathbf{y}_i$, respectively.

We generate synthetic data using a finite number of mixture components in the DPMPM.
Paraphrasing from \citet{SiReiter2013}, the
finite DPMPM assumes that each individual $i$ belongs to exactly one
of $F < \infty$ latent
classes; see \citet{SiReiter2013} for advice on
determining $F$.  For $i=1, \dots, n$, let $\eta_i \in \{1, \dots,
F\}$ indicate
the class of individual $i$, and let $\pi_f = \mbox{Pr}(\eta_i = f)$. We assume that $\mathbf{\pi} = (\pi_1, \dots, \pi_F)$
is the same for all individuals. Within any class, each of the $p$ variables independently follows a class-specific multinomial
distribution, so that individuals in the same latent class have the same cell probabilities. For any value $c \in \{1,\dots, d_{k}\}$,
let $\phi_{fc}^{(k)} = \mbox{Pr}(Y_{ik}=c \mid \eta_{i}=f)$
be the probability of $Y_{ik}=c$, given that individual $i$ is in
class $f$. Let $\mathbf{\phi} = \{\phi_{fc}^{(k)}: c=1,\dots,d_k, k=1,\dots,p, f=1,\dots, F\}$ be the collection of all $\phi_{fc}^{(k)}$.
The finite mixture model can be expressed as
\begin{eqnarray}
	\label{lca-x} Y_{ik} \mid \eta_i, \phi &\overset{ind}{\sim}& \textrm{Multinomial}(\phi_{\eta_i1}^{(k)}, \dots,
		\phi_{\eta_i d_k}^{(k)}) \,\,\,\, \textrm{for all } i,k,\\
	\label{lca-eta} \eta_i \mid \pi &\sim& \textrm{Multinomial}(\pi_1, \dots, \pi_F) \,\,\,\, \textrm{for all } i,
\end{eqnarray}
where each multinomial distribution has a sample size equal to one and the number of levels is implied by the dimension
of the corresponding probability vector.

For prior distributions on $\pi$ and $\phi$, we use the truncated
stick-breaking representation of \citet{Sethuraman1994}. We have
\begin{eqnarray}
\label{modelprior}
	\pi_f &=& V_{f}\prod_{l<f}(1-V_{l}) \,\,\,\, \textrm{for } f=1, \dots, F,\\
	V_{f}&\overset{iid}{\sim}& \textrm{Beta} (1,\alpha) \,\,\,\, \textrm{for } f=1, \dots, F-1, \,\,\,\, V_F=1,\\
	\alpha &\sim& \textrm{Gamma}(a_{\alpha}, b_{\alpha}),\\
	\mathbf{\phi}_{f}^{(k)}=(\phi_{f1}^{(k)},\dots,\phi_{fd_k}^{(k)})&\sim& \textrm{Dirichlet} (a_{k1}, \dots,a_{kd_{k}}).
\end{eqnarray}
We set $a_{k1} = \dots = a_{kd_{k}} = 1$ for all $k$ to correspond to
uniform distributions. Following
\citet{dunsonxing} and \citet{SiReiter2013}, we set $(a_{\alpha} =.25,  b_{\alpha}=.25)$,
which represents a small prior sample size and hence a vague specification for the Gamma distribution.
In practice, we find these specifications allow the data to dominate
the prior distribution. We estimate the posterior distribution of
all parameters using a blocked Gibbs sampler \citep{ishwaranjames}.
%

We illustrate how to generate one synthetic dataset, assuming that only the $p$th variable, $Y_p$, should be synthesized. Thus, $\mathbf{\phi}^{(p)}$ contains all multinomial probabilities associated with $Y_p$. To generate one partially synthetic dataset of size $n$, we first sample a value of the parameters $(\alpha, \mathbf{V}, \mathbf{\pi}, \mathbf{\phi}^{(p)})$ from their respective posterior distributions. Using the drawn value of $\pi$, we sample values of $(\eta_1, \dots, \eta_{n})$ independently from (\ref{lca-eta}). Using the sampled $\phi^{(p)}$, for each sampled $\eta_i$ ($i=1,\dots,n$),  we then sample the $i$th synthetic record, $\mathbf{y}_{i}^*=(y_{i1}, \dots, y_{ip-1}, y_{ip}^*)$, from a multinomial distribution with probabilities $\phi_{\eta_i}^{(p)}$ for $Y_p$.  The synthesis can be conveniently implemented inside the blocked Gibbs sampler -- after each Gibbs updating step, we simply sample and save draws of $\mathbf{y}_{i}^*$ for all $n$ records.
To create $m>1$ synthetic datasets, one repeats this process $m$ times, using approximately independent draws of parameters. Approximately independent draws can be obtained by using iterations that are far apart in the
estimated MCMC chain.

If more than one variable is to be synthesized, each synthetic variable can be generated independently at the desired iterations. Suppose there are $r$ ($r\leq p$) variables to be synthesized, and let $Y_{l}$ ($l = 1, \dots, r$) represent the $l$th variable to be synthesized. Assume that the variables in the dataset are ordered so that the $p-r$ variables that remain unaltered appear first. After sampling $\eta_i$, we can sample the $i$th synthetic record, $\mathbf{y}_{i}^*=(y_{i1}, \dots, y_{ip-r},y_{ip-r+1}^*,\dots, y_{ip}^*)$, from corresponding multinomial distributions with probabilities $\mathbf{\phi}_{\eta_i}^{(l)}$ for $Y_l$.
Note that the synthesis order does not matter, because each variable independently follows a multinomial distribution given the latent class assignment.

\section{The CART Synthesizers}\label{CART_detail}
The following description of the CART synthesizers borrows heavily from \citet{DrechslerReiter2011}. Those interested are referred to this paper for more details on CART synthesizers and other machine learning approaches for data synthesis.

First the agency fits the tree of $Y_p$ conditioning on all other variables in the dataset so that each leaf contains
at least $d$ records; call this tree $\mathcal{Y}^{(p)}$. In general, we have
found that using the default specification, which varies between $d=5$ and $d=7$ depending on the software,
 provides sufficient accuracy and reasonably fast
running time. We cease splitting any particular leaf when the ``impurity'' in that leaf is
less than some agency-specified threshold, or when we cannot ensure at least $d$ records
in each child leaf. The ``impurity'' basically measures the heterogeneity of the outcome variable in each leaf. For continuous variables, the variance in each leaf is commonly used as an impurity measure. For categorical variables, the Gini coefficient or the entropy are typically employed. We use the Gini coefficient in our application since it is recommended for categorical variables with more than two categories \citep{Berk2008}. See \citet{Berk2008}, for example, for a more detailed discussion of impurity measures.
For all records $i=1,...,n$ in the original data, we trace down the branches of
$\mathcal{Y}^{(p)}$ until we find that record's terminal leaf.
Let $L_{w}$ be the $w$th terminal leaf in $\mathcal{Y}^{(p)}$,
and let $Y^{(p)}_{L_{w}}$ be the $n_{L_{w}}$
values of $Y_p$ in leaf $L_{w}$. For all records whose terminal leaf
is $L_{w}$, we generate replacement values of $Y_{pi}$ by drawing from $Y^{(p)}_{L_{w}}$
using the Bayesian bootstrap (Rubin, 1981). Repeating the
Bayesian bootstrap for each leaf of $\mathcal{Y}^{(p)}$ provides one synthetic dataset. We repeat this process $m$ times to
generate $m$ datasets with synthetic values for $Y_p$.

If more than one variable should be synthesized, a sequential regression multivariate imputation approach (SRMI, \citet{Raghu2001}) can be used. In such cases, the agencies can proceed as follows for an arbitrary ordering of the variables. Let $Y_{l}$ represent the $l$th variable in the synthesis order and let $Y_{0}$ be all
variables with no values replaced.
\begin{enumerate}
\item  Run the CART algorithm to regress $Y_{1}$ on $Y_{0}$ only. Replace
  $Y_{1}$ by synthetic values using the corresponding synthesizer for $Y_{1}$. Let $Y_{1}^*$
  be the replaced values of $Y_{1}$.
\item  Run the algorithm to regress $Y_{2}$ on $(Y_{0}, Y_{1})$ only.
Replace $Y_{2}$ with synthetic values using the corresponding synthesizer
  for $Y_{2}$. Use the values of
  $Y_{1}^*$ and $Y_{0}$ for predicting new values for $Y_{2}$. Let $Y_{2}^*$
  be the replaced values of $Y_{2}$.
\item For each $l$ where $l=3, \dots, r$, run the algorithm to regress
  $Y_{l}$ on $(Y_{0}, Y_{1}, \dots,$ $Y_{l-1})$. Replace each
  $Y_{l}$ using the appropriate synthesizer based on the values in $(Y_0, Y_{1}^*,$
  $Y_{2}^*, \dots, Y_{l-1}^*)$.
\end{enumerate}
The result is one synthetic dataset. These three steps are repeated
for each of the $m$ synthetic datasets, and these datasets are
released to the public.

\section{Methodology for Estimating the Risk of Disclosure}\label{risk_detail}
The description of the methodology follows the description given in \citet{DrechslerReiter2010}. Suppose the intruder has a vector of information, $\mathbf{t}$, on a particular
target unit in the population $\mathbf{P}$.
Let $t_0$ be the unique identifier of the target,
and let $P_{i0}$ be the unique identifier (not released) 
for record $i$ in $\mathbf{d}_{syn}$, where $\mathbf{d}_{syn}$ denotes the synthetic data and $i=1,\dots,n$.
Let $\mathcal{S}$ be any information released about the simulation models.

The intruder's goal is to match unit $i$ in $\mathbf{d}_{syn}$
to the target when  $P_{i0}=t_0$. Let $J$ be a
random variable that equals $i$ when $P_{i0}= t_0$ for
$i \in \mathbf{d}_{syn}$. The intruder thus seeks to calculate
$Pr(J=i |\mathbf{t},  \mathbf{d}_{syn}, \mathcal{S})$
for $i=1,\dots, n$. Because the intruder does not know the
actual values in $Y^*$, she should integrate over
its possible values when computing the match probabilities. Hence,
for each record we compute
\begin{eqnarray}\label{eq:risk}
Pr(J=i |\mathbf{t}, \mathbf{d}_{syn}, \mathcal{S}) = \int Pr(J=i | \mathbf{t},
\mathbf{d}_{syn}, Y^*, \mathcal{S}) Pr(Y^*|\mathbf{t}, \mathbf{d}_{syn}, \mathcal{S}) d Y^*.
\end{eqnarray}
This construction suggests a Monte Carlo approach to estimating
each $Pr(J=i |\mathbf{t}, \mathbf{d}_{syn}, \mathcal{S})$. First, sample a
value of $Y^*$ from $Pr(Y^*|\mathbf{t},
\mathbf{d}_{syn}, \mathcal{S})$. Let $Y_{new}$ represent one set of
simulated values. Second, compute
$Pr(J=i|\mathbf{t}, \mathbf{d}_{syn}, Y^*=Y_{new}, \mathcal{S})$
using exact matching, assuming $Y_{new}$ are
collected values. This two-step process is iterated $h$
times, where ideally $h$ is large, and (\ref{eq:risk}) is estimated as
the average of the resultant $h$ values of $Pr(J=i|\mathbf{t},
\mathbf{d}_{syn}, Y^*=Y_{new}, \mathcal{S})$. When $\mathcal{S}$ has no
information,  the intruder treats the simulated values as
plausible draws of $Y^*$.

Following \citet{Reiter2005CART}, we quantify disclosure risk with summaries of
these identification probabilities. It is reasonable to assume that the intruder
selects the record $i$ with the highest value
of $Pr(J=i|\mathbf{t}, \mathbf{d}_{syn}, \mathcal{S})$ as a match for $\mathbf{t}$.
We consider two risk measures: the expected match risk and the true match rate. Let $c_i$ be the number of records with the highest match probability for the target $\mathbf{t_i}$; let $I_i=1$ if the true match is among the $c_i$ units, and $I_i=0$ otherwise. The expected match risk equals $\sum I_i/c_i$. When $I_i=1$ and $c_i>1$, the contribution of unit $i$ to the expected match risk reflects the intruder randomly guessing at the correct match from the $c_i$ candidates. Let $K_i=1$ when $c_iI_i=1$, and $K_i=0$ otherwise, and let $N$ denote the total number of target records. The true match rate equals $\sum{K_i}/N$, which is the percentage of true unique matches among the target records. 


\end{document}


\begin{flushleft}
  {\bf Online supplement to the paper: Synthesizing geocodes to facilitate access to detailed geographical information in large-scale administrative data}
  \vspace{1.0cm}

  J{\"o}rg Drechsler$^*$, Jingchen Hu$^{**}$

{\small
\begin{description}
\item $^* \;$ Institute for Employment Research, Regensburger Str. 104, 90478 Nuremberg, Germany, joerg.drechsler@iab.de
\item $^{**}\;$ Department of Mathematics and Statistics, Vassar College, Poughkeepsie, NY 12604, USA, jihu@vassar.edu
\end{description}
}
\end{flushleft}
\vspace{1cm}

This online supplement consists of four sections. In the first section, we provide details regarding the dataset used in the main paper, the Integrated Employment Biographies (IEB). In Section 2, we provide some convergence diagnostics for the DPMPM synthesizer. In Section 3, we present detailed results regarding the impacts on risk and analytical validity if additional measures beyond synthesizing the geocodes are taken to protect the data further. Specifically, we look at two possible strategies: aggregating the geographical information to a higher level (Section \ref{geo_aggregate}) or synthesizing more variables (Section \ref{synmore}). We only present the results for the categorical CART synthesizer in this supplement since the risk levels are arguably already acceptable for the other two synthesizers. However, we ran all simulations for all synthesizers and verified that the general findings regarding the relative performance of the three different synthesizers remained the same. Detailed results for the other synthesizers can be obtained from the authors upon request.

Finally in Section \ref{cp}, we evaluate the influence of the complexity parameter $cp$ -- the most important tuning parameter for CART models -- on the analytical validity and risk of disclosure of the generated synthetic data. We again focus on the categorical CART synthesizer, as it outperformed the other synthesizers in terms of analytical validity. We expect to see similar patterns for the continous CART synthesizer.

\section{The Integrated Employment Biographies}\label{IEB_detail}
The Integrated Employment Biographies (IEB) integrate five different sources of information collected by the Federal Employment Agency through different administrative procedures: the Employment History (BeH), the Benefit Recipient History (LeH), the Participants-in-Measures History (MTH), the Unemployment Benefit II Recipient History (LHG), and the Jobseeker History (ASU). The different data sources are integrated for and by the Institute for Employment Research.

\begin{figure}[!t]
\includegraphics[width=\textwidth]{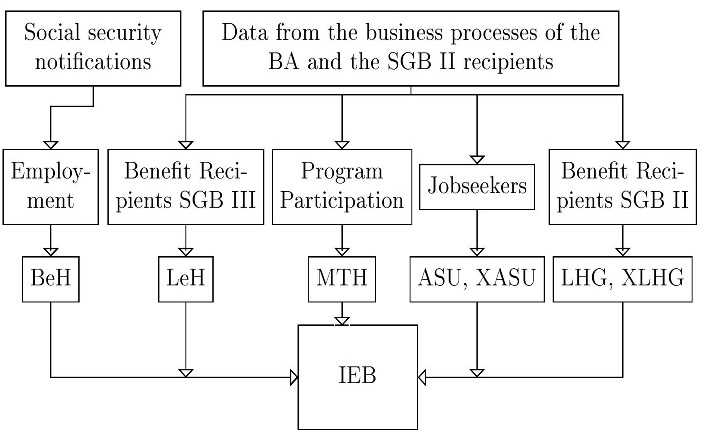}
\caption{Overview of the different administrative data sources integrated into the IEB.}
\label{fig:IEB1}
\end{figure}

Figure \ref{fig:IEB1} provides an overview of the business processes that generate IEB data. BeH information is provided for every employee covered by social security. 
Individuals who did not enter the labor market and self-employed were excluded, since these groups are not subject to mandatory social security contributions. 
LeH and (X)LHG data are generated for individuals receiving benefits in accordance with the German Social Code Books (SGB) II \citep{SGBII} and III \citep{SGBIII} (SGB II regulates the welfare benefits for employable jobseekers in need, and SGB III regulates the employment promotion, and unemployment insurance in particular). MTH and (X)ASU data are generated for individuals who were registered as jobseekers with the Federal Employment Agency or who participated in an employment or training program. 
Information available in the IEB includes, for example:
\begin{description}
\item[-] Dates when each spell of employment begins and ends
\item[-] Date of birth, gender, and nationality
\item[-] Education and health status
\item[-] Monthly wage
\item[-] Employment status, and employment status prior to job search
\item[-] Industry
\item[-] Place of work and place of residence on the ZIP code level
\item[-] First day in employment and number of days working for the current employer
\end{description}
 Establishment-level information is provided in the IEB by aggregating individual-level information to the establishment level. We refer to \cite{JacobebinghausSeth2010} for a detailed description of the different data sources and the IEB.

External researchers interested in working with the data can apply for access to a 2\% sample from the IEB containing a limited set of non-sensitive variables and county-level information as the lowest level of geographical detail. Counties with fewer than 100,000 observations in the full IEB are collapsed. Several additional steps, such as aggregating some variables or dropping employment information outside a given age range, are taken to ensure a sufficient level of confidentiality protection (see Section 3.4 in \citet{ganzer2017} for further details regarding the anonymization measures). We also note that these data can only be accessed by scientific researchers (not the general public) after a positive review of their research proposal, which must demonstrate an important contribution to labor market research and also include convincing arguments as to why the research cannot be addressed with other data sources. Approved researchers can decide whether they need access to the detailed version of the data, which can only be accessed at the research data center of the IAB, or if a very limited set of variables -- which can be disseminated as scientific use files -- is sufficient to answer their research question.

\section{Details regarding the specification of the MCMC sampler and convergence diagnostics for the DPMPM synthesizer} \label{MCMC_detail}

In each cluster, we ran the MCMC sampler for the DPMPM synthesizer for 10,000 iterations, treating the first 5,000 iterations as burn-in and storing only every 10th iteration to reduce the correlation between the successive draws. The 5,000 iterations from the burn-in period (i.e., the first 5,000 iterations at the beginning of the MCMC chain) are not used, to ensure that the MCMC sampler converged before any parameters are drawn for synthesis. To monitor convergence and autocorrelation, we focused on the parameter $\alpha$, which is not subject to label switching. We used the Geweke, and Heidelberger and Welch's diagnostics, and inspected the autocorrelation function to evaluate the behavior of the MCMC sampler for this parameter. We reran the MCMC sampler and stored only every 50th iteration for those clusters which expressed high correlation among successive draws. We did not find any problematic cases based on our evaluation criteria after this extra step. We set the maximum number of allowed latent classes to $F=100$.  The posterior mean of the number of occupied classes is 57.8, with a 95\% central interval ranging from 51 to 64 and a maximum of 73. 

\section{Additional methods for protecting the data}
\subsection{Aggregating the geocoding information}\label{geo_aggregate}
Aggregation of detailed geographical information is the classical approach that most statistical agencies choose when disseminating data to the public. In this section, we evaluate the impact of this strategy if used in combination with data synthesis, that is, we assume that the original data are aggregated first before the aggregated information is synthesized in a second step. For the risk of disclosure, we expect two counterbalancing effects from this strategy. On the one hand, disclosure risks will generally decrease, since the number of individuals that share the same geographical information rises with increasing aggregation, and thus it will be more difficult for the intruder to identify individuals uniquely using this information. On the other hand, aggregating the geographical information implies that the number of levels for geography will decrease. Since the selected synthesizer treats geography as a categorical variable, we expect that the fit of the CART model will increase, and more terminal nodes of the tree will contain only one geography, which, in turn, implies that the synthetic geography will match the true geography more often, thus offering less protection.

The impacts on analytical validity are more difficult to anticipate. The improved fit of the CART model will generally imply higher analytical validity, but analysis on a very detailed level will obviously no longer be possible. Moreover, even on the aggregated level, the boundaries of the geographic area the analyst is interested in will not necessarily coincide with the boundaries based on the aggregation level chosen by the agency.

We evaluate the impacts on validity and risk using three different aggregation levels: aggregation to 10, 100, and 1,000 square meter grids. The aggregated geocodes are obtained by flooring the latitude and longitude values according to the selected aggregation level. To measure the analytical validity, we use the \emph{UL} measure defined in the main paper. Since the grid levels do not necessarily match with the ZIP code levels, we assume that the analyst would simply assign the ZIP code that is closest to the released geocode. This strategy could be improved by sampling from all ZIP codes that fall in the same grid. However, we found that even for the 1,000 square meter grids, less than 7.5\% of the grids contained more than one ZIP code ($<0.25\%$ for the 100 square meter girds, and only 4 out of over 1.6 million grids for the 10 square meter grids). Thus, we do not expect improving the ZIP code assignment to have a major impact on the results.

Results for the different aggregation levels are presented in Figure \ref{fig:dr_aggr}. 
For the risk measures, we only present the results assuming the intruder would pick the level of aggregation that leads to the maximum risk. We also include the utility and risk measures for the exact geocodes from the main text for comparison.

\begin{figure}[t]
\begin{center}
  \includegraphics[scale=0.55,trim=0cm 0cm 0cm 0cm]{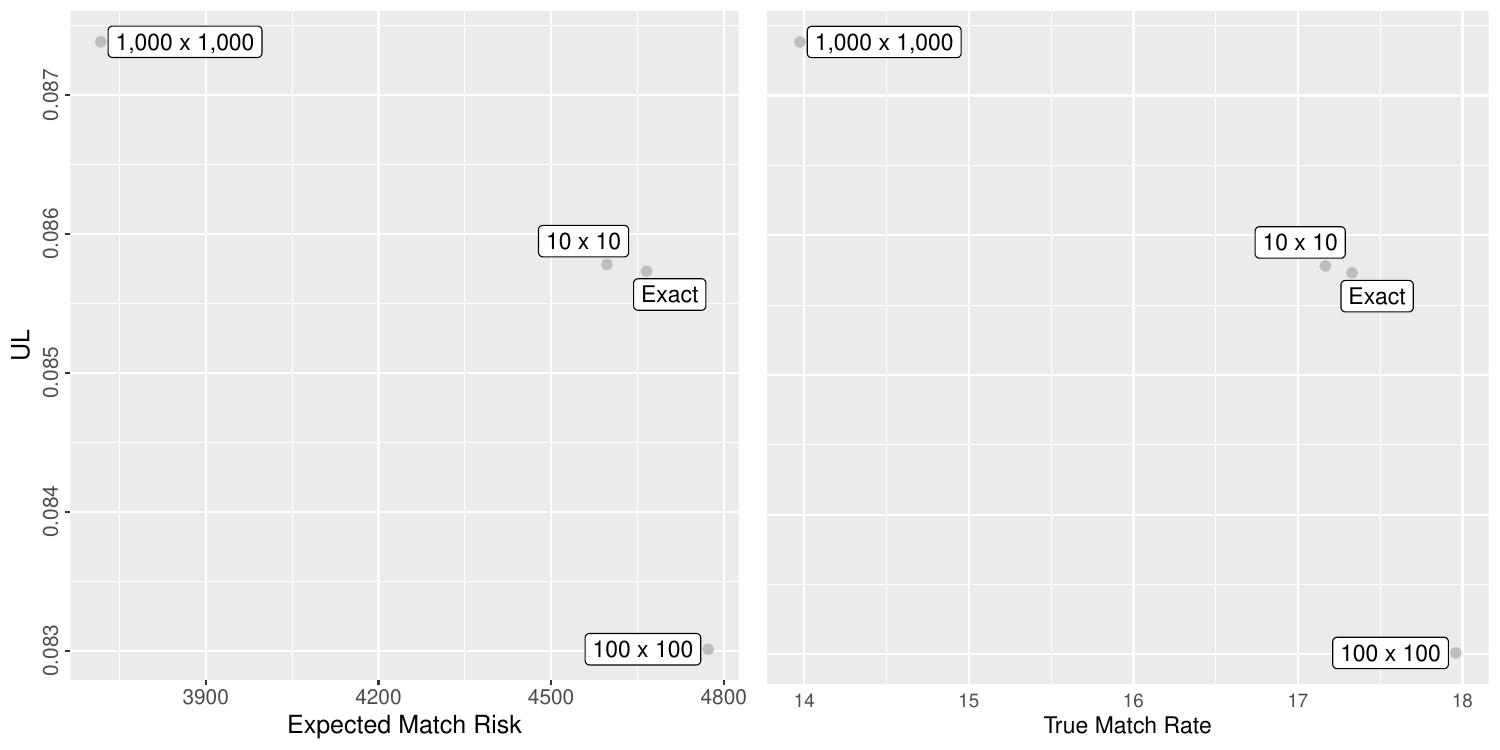}\label{fig:dr_aggr}\\
  \caption{Global utility loss (\emph{UL}), expected match risk, and true match rate (in \%) for various aggregation levels.}\label{fig:dr_aggr}
  \end{center}
\end{figure}


Figure \ref{fig:dr_aggr} indicates that the utility and risk remain more or less constant when moving from the exact geocodes to the 10 square meter aggregation. The analytical validity is highest if the geocodes are aggregated to 100 square meter grids, but the improvements in validity come at the price of an increased risk. Only for the 1,000 square meter grids do we see a decrease in the risk measures compared to the exact geocodes. 
However, the risk levels are still considerably higher than the risk levels for the original data without geocodes (the expected match risk and true match rate are 1821.16 and 2.56\%, respectively for these data). Thus, the agency would have to select even larger aggregation levels to protect the data to a sufficient extent. However, the \emph{UL} measure indicates that the analytical validity starts decreasing once the data are aggregated beyond the 100 square meters grids. We emphasize again that any analysis based on a finer level of detail than the ZIP code would likely be even more affected by this aggregation step.

\subsection{Synthesizing more variables}\label{synmore}
\begin{figure}[t]
\begin{center}
  \includegraphics[scale=0.55,trim=0cm 0cm 0cm 0cm]{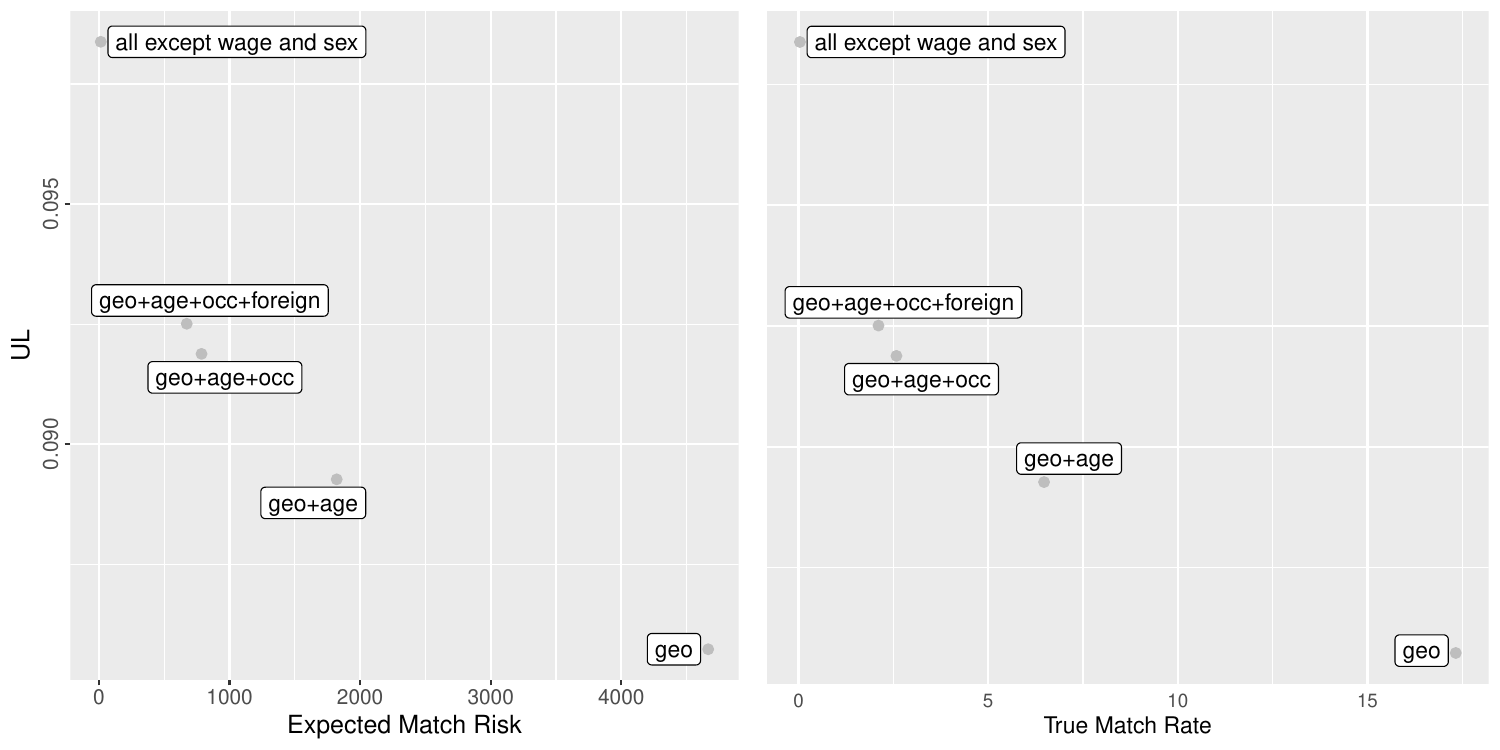}\\
  \caption{Global utility loss (UL), expected match risk, and true match rate (in \%) for various amounts of synthesis.}\label{fig:synmore}
\end{center}
\end{figure}

As an alternative approach to increasing the protection offered by the categorical CART synthesizer, we evaluate synthesizing more variables in this section. We start by synthesizing the age variable in addition to the geocode variable. Then we successively add occupation and foreign status to the list of variables that are synthesized. Finally, we evaluate a scenario in which all variables except sex and wage are synthesized. Note that in this scenario some variables are synthesized which, according to our risk scenario, are not known by the intruder. Thus, synthesizing these variables will only have negative impacts on our utility measures but will not decrease the risks. This setting can be seen as a conservative approach in which the agency expects that additional variables might be available to the intruder in the future even if they are not available now. We note that the sensitive wage information is not synthesized in any of the scenarios. Not synthesizing wage implies that the synthesis is only conducted to prevent reidentification. Obviously, the synthesis could also be used to protect the sensitive information in the data directly. However, different risk measures would be necessary to evaluate the success of this strategy. Our risk measures focus on quantifying the risks of reidentification. The results for these measures would not change if the wage information were also synthesized. Thus, we refrain from synthesizing the wage information in this evaluation.
Results for the different amounts of synthesis are presented in Figure \ref{fig:synmore}.

As expected, both analytical validity and disclosure risk decrease with increasing amounts of synthesis. However, the risks decrease more substantially. For example, if age is synthesized in addition to the geocode, the \emph{UL} measure increases by 4.1\%, while the expected risk and the true match rate decrease by 61\% and 63\%, respectively. The loss in validity is generally minor, and never greater than 8\% except for the last scenario, in which almost all variables are synthesized, leading to negligible risk. However, even in this scenario, the \emph{UL} measure is substantially lower than for the other two synthesizers, even if we assume -- as in the main text -- that these synthesizers would only be used to synthesize the geocode ($9.84\cdot10^{-2}$ compared to $10.95\cdot10^{-2}$ for the continuous CART synthesizer, and $14.94\cdot10^{-2}$ for the DPMPM synthesizer if only the geocode is synthesized). We also note that the analytical validity if age, occupation, and foreign status are synthesized is comparable to the analytical validity if all geocodes are aggregated to 1,000 square meter grids. However, the risks of disclosure are substantially smaller. Finally, the expected risks for the scenario in which only age and geocodes are synthesized are comparable to the risks under the current access model, that is, without geocodes but with no alteration for any of the other variables (1820.46 versus 1821.16). The percentage of correct unique matches among the target records (the true rate) is slightly larger (6.47\% versus 2.56\%), but arguably, there will be many declared unique matches in the synthetic data that are wrong. Given that the intruder would know under the current access strategy that all unique matches were correct matches, we feel that the overall risks would be lower if age and geocode were synthesized. Thus, from a risk perspective, it seems justified for the IAB to provide access to a 2\% sample of the data in which only the geocodes and the age variable are synthesized, but the geocoding information is not aggregated. In this scenario, the estimated risks are lower than the risks for a dataset that can already be accessed on the premises of the IAB. Alternatively, the IAB could opt for synthesizing more variables and releasing a larger sample instead.

\section{Varying the complexity parameter of the CART model}
\label{cp}
The complexity parameter ($cp$) is an important tuning parameter for CART models. Any split that does not decrease the overall lack of fit by a factor of $cp$ is not attempted. Thus, if at a specific point of the tree-growing process all further splits of the data cannot improve the purity of the resulting nodes beyond the level required by the complexity parameter, the final size of the tree is established. Consequently, setting $cp$ to a large value will result in small trees, whereas small values of $cp$ will typically result in larger trees. \citet{DrechslerReiter2011} recommend that for data synthesis the value of the complexity parameter should be set to smaller values than the standard recommendations found in the literature. In this section, we evaluate the impacts of changing the $cp$ value for the categorical CART synthesizer. Specifically, we look at the following $cp$ values: $cp=\{5\cdot10^{-4},4\cdot10^{-4},\ldots,1\cdot10^{-4},0.9\cdot10^{-4},0.8\cdot10^{-4},\ldots,0.5\cdot10^{-4}\}$.

\begin{figure}[t]
\begin{center}
  \includegraphics[scale=0.6,trim=0cm 0cm 0cm 0cm]{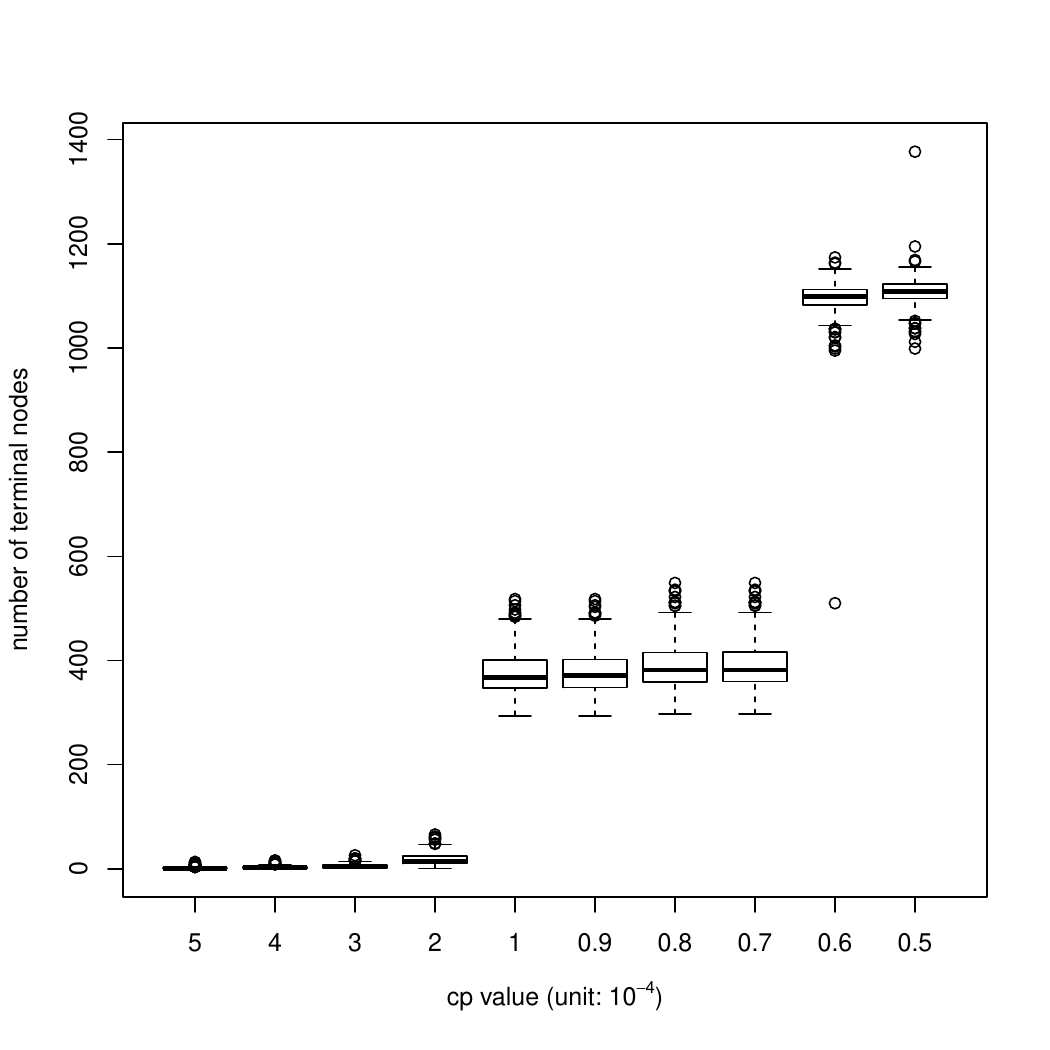}
  \caption{Boxplots for the number of terminal nodes for various levels of the complexity parameter ($cp$).}\label{fig:nodes}
  \end{center}
\end{figure}
Figure \ref{fig:nodes} provides boxplots of the number of terminal nodes across the 222 clusters for the different $cp$ values (remember that independent trees are grown within each cluster). The figure already reveals the main disadvantage of using the $cp$ value as a tuning parameter. The tree sizes remain relatively constant over a whole range of $cp$ values, with sudden jumps at certain values. In practice, it would be cumbersome to identify the values of the parameter which lead to substantial changes in the tree size.

\begin{figure}[t]
\begin{center}
  \includegraphics[scale=0.55,trim=0cm 0cm 0cm 0cm]{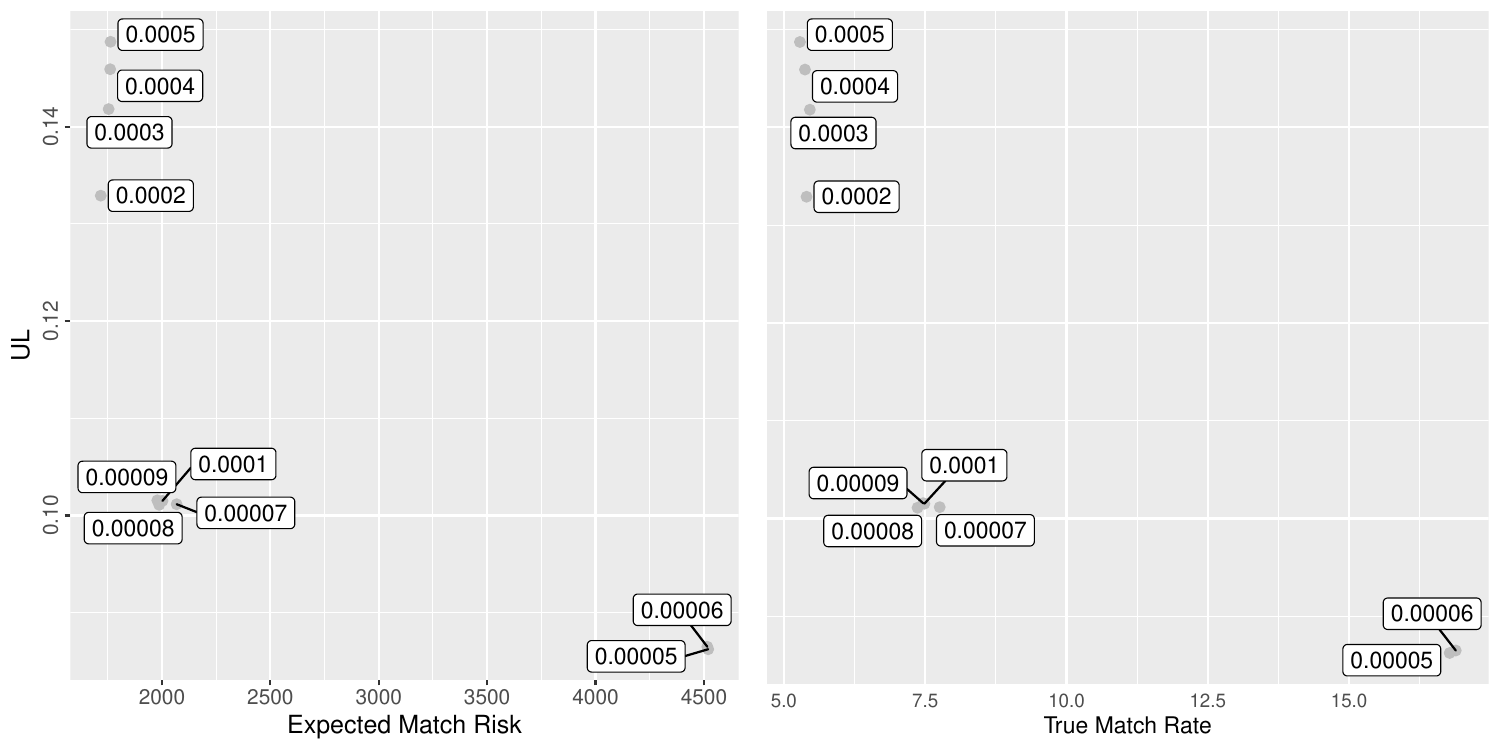}
  \caption{Global utility loss (\emph{UL}), expected match risk, and true match rate (in \%) for various values of the complexity parameter.}\label{fig:ru_map}
  \end{center}
\end{figure}

Figure \ref{fig:ru_map} presents risk-utility maps for the different $cp$ values. For the risk measures, we assume that the intruder would always use the level of aggregation for the geocoding information which leads to the highest risk levels among the following grid sizes: 1, 10, 50, 100, 500, 1,000, 5,000, 10,000, and 20,000 square meter grids. We acknowledge that this will lead to conservative estimates for the true risks, as the intruder will not know in practice which level of aggregation results in the highest probability of finding a correct match.

The results in Figure \ref{fig:ru_map} are consistent for both risk measures. Changing the $cp$ value between $5\cdot10^{-4}$ and $2\cdot10^{-4}$ has almost no effect on the level of disclosure risk. However, it slightly improves the analytical validity of the data as measured by the \emph{UL} measure. Analytical validity and disclosure risk increase substantially when the $cp$ value is further reduced to $1\cdot10^{-4}$. In the range between $1\cdot10^{-4}$ and $0.7\cdot10^{-4}$, both risk and validity remain constant, but a further reduction to $0.6\cdot10^{-4}$ again leads to substantial increases in both dimensions. Further reducing the $cp$ value has no effect on risk or utility.

\begin{figure}[t]
\begin{center}
  \includegraphics[scale=0.6,trim=0cm 0cm 0cm 0cm]{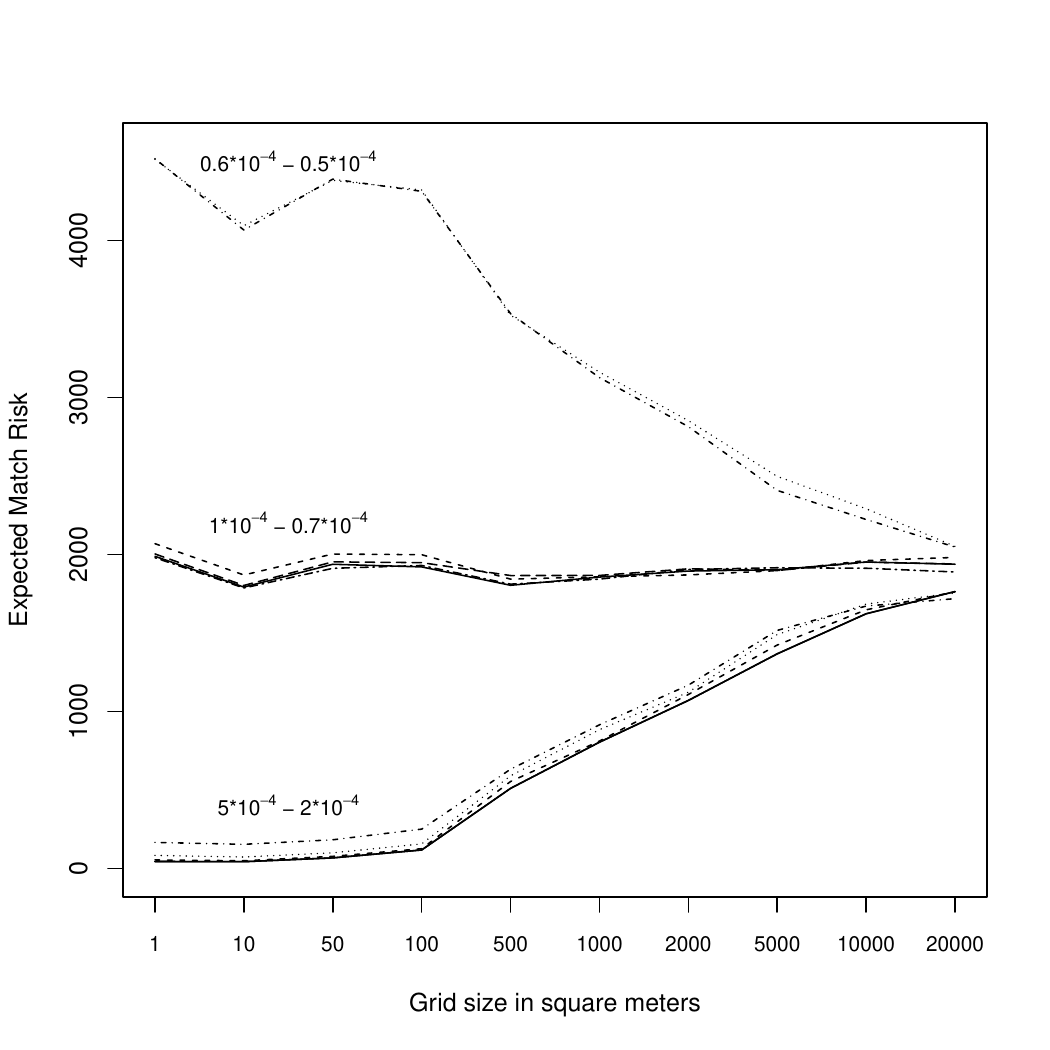}
  \caption{Relationship between expected match risk and level of aggregation of the geocoding information for various levels of the complexity parameter. Each line represents one $cp$ value. To avoid clutter, we only give ranges for the $cp$ values which show similar behavior instead of labeling each line separately.}\label{fig:risk_vs_agg}
  \end{center}
\end{figure}
We also evaluated whether the optimal level of aggregation that an intruder should pick depends on the value of the complexity parameter. Results for the expected match risk are presented in Figure \ref{fig:risk_vs_agg}. Each line represents a specific $cp$ value. We refrain from labeling each line to avoid clutter. Instead, we only label the ranges of the $cp$ value for which different patterns emerge. Between $5\cdot10^{-4}$ and $2\cdot10^{-4}$ the risk increases monotonically with increasing levels of aggregation. This is not surprising, as the geocodes in the synthetic data are unlikely to be close to the geocodes from the original data if very small trees are grown. Thus, increasing the grid size in which a record will be declared a match will generally increase the probability of the correct match being included among the declared matches. In the range between $1\cdot10^{-4}$ and $0.7\cdot10^{-4}$, changing the grid size has almost no impact on the risk. This means that the increasing probability of the correct match being among the declared matches is counterbalanced by the increasing probability of finding more than one matching record, which reduces the probability of the correct match being identified. Finally, for $cp$ values of $0.6\cdot10^{-4}$ or lower, the risk mostly decreases with increasing levels of aggregation. In these scenarios, the generated synthetic geocode is close to the original geocode in a sufficient number of cases for the increasing probability of the correct match being among the declared matches to be outweighed by the increasing probability of finding more records that match on all matching criteria.

We also evaluated $cp$ values outside the range reported in Figures \ref{fig:nodes} to \ref{fig:risk_vs_agg}. For $cp$ values of 0.005 or larger, the CART models never split the data in any of the clusters, and the analytical validity and risk results are comparable to the results for $cp=5\cdot10^{-4}$. When reducing the $cp$ value from $cp=0.5\cdot10^{-4}$ to $cp=1\cdot10^{-6}$, the average number of final nodes slightly increases from 1107 to 1139. However, the impacts on analytical validity and risk are negligible.

Summarizing our evaluations, we find that both utility and risk tend to increase with decreasing values of the complexity parameter, as expected. However, the relationship is not smooth. Both dimensions remain constant for certain ranges of the $cp$ value, but sometimes change substantially as a result of small changes in the $cp$ value. The reason for this is that the tree size also remains relatively stable within certain ranges of the $cp$ value, but can sometimes increase substantially as a result of small changes in the $cp$ value. Overall, this means that the $cp$ value cannot be used to fine-tune the risk-utility trade-off for synthetic data flexibly. If the data disseminating agency is not satisfied with the level of risk achieved with CART models setup with a small $cp$ value, we suggest using some of the other measures discussed in this online supplement to increase the level of protection further.

\bibliographystyle{natbib}
\bibliography{geo_synth_bib}